\documentclass[preprint,aps,nofootinbib]{revtex4-1}	
\usepackage{amsmath}
\usepackage{amsfonts}
\usepackage{amssymb}
\usepackage{color}
\usepackage{bm}		% bolded Greek letters
\usepackage{graphicx}
\usepackage{feynmp}
\usepackage{slashed}	% for slashed characters in math mode
\usepackage{bbm}		% for \mathbbm{1} (unit matrix)
\usepackage{xspace}	% For spacing after commands
\usepackage{empheq}
\usepackage[toc,page]{appendix}
\usepackage{hyperref}	% Has to be at the end.

\providecommand{\be}[1]{\begin{equation}{#1}\end{equation}}  
\providecommand{\ba}[1]{\begin{align}{#1}\end{align}}
\newcommand{\nl}{\nonumber \\ }

\providecommand{\mean}[1]{\langle #1 \rangle}

\providecommand{\dt}[1]{\frac{\mathrm{d} #1}{\mathrm{d} t}} % derivative with respective to \log Q
\providecommand{\dnt}[2]{\frac{\mathrm{d}^#2 #1}{\mathrm{d} t^#2}} % nth-derivative with respective to \log Q

\def\GeV{\text{ GeV}} % GeV
\def\TeV{\text{ TeV}} % GeV
\def\Nfl{N_{\text{fl}}} % number of flavors
\providecommand{\wt[1]}{\widetilde{#1}} % Overtilde, DR-bar scheme, appendix B
\providecommand{\wh[1]}{\widehat{#1}} % Normalization by M_S
\def\DRbar{\overline{\text{DR}}} % DR-bar scheme
\def\MSbar{\overline{\text{MS}}} % MS-bar scheme
\def\mb{\overline{m}} % MS-bar running top-quark mass
\def\MSUSY{M_{\text{SUSY}}} % SUSY mass scale, arithmetic average of squark masses
\def\tt{\tilde{t}} % stop or squark
\def\muMS{\hat{\mu}} % OS Higgs SUSY mu normalized by M_S
\def\XtMS{\widehat{X}_t}% MS-bar stop mixing parameter normalized by M_S 
\def\tb{t_\beta} % Higgs vev angle
\def\cb{c_\beta}
\def\sb{s_\beta}

\def\a{\alpha}
\def\b{\beta}
\def\c{\chi}
\def\d{\delta}

\def\k{\kappa}
\def\l{\lambda}
\def\m{\mu}

\def\q{\theta}

\def\t{\tau}

\def\D{\Delta}
\def\F{\Phi}

\begin{document}
%\begin{titlepage}

\begin{flushright}
SCIPP 13/15\\
EFI Preprint 13-29 \\
\end{flushright}

\vspace{0.3cm}
\begin{center}
\Large\bf 
Precise Estimates of the Higgs Mass in Heavy Supersymmetry
\end{center}

\vspace{0.5cm}
\begin{center}
{\sc Patrick Draper$^{(a)}$, Gabriel Lee$^{(b)}$, and Carlos E.~M.~Wagner$^{(b,c,d)}$ } \\
\vspace{0.2cm}
{\it 
$^{(a)}$ Santa Cruz Institute for Particle Physics, Santa Cruz, California 95064, USA
}\\ 
\vspace{0.3cm}
{\it
$^{(b)}$ Enrico Fermi Institute and Department of Physics, \\
University of Chicago, Chicago, Illinois 60637, USA
}\\
\vspace{0.3cm}
{\it
$^{(c)}$ Kavli Institute for Cosmological Physics, \\
University of Chicago, Chicago, Illinois 60637, USA
}\\
\vspace{0.3cm}
{\it 
$^{(d)}$ HEP Division, Argonne National Laboratory, \\
9700 Cass Avenue, Argonne, Illinois 60439, USA
}
\end{center}
\vspace{0.5cm}

\begin{abstract}
In supersymmetric models, very heavy stop squarks introduce large logarithms into the computation of the Higgs boson mass. Although it has long been known that in simple cases these logs can be resummed using effective field theory techniques, it is technically easier to use fixed-order formulas, and many public codes implement the latter. We calculate three- and four-loop next-to-next-to-leading-log corrections to the Higgs mass and compare the fixed order formulas numerically to the resummation results in order to estimate the range of supersymmetry scales where the fixed-order results are reliable. We find that the four-loop result may be accurate up to a few tens of TeV. We confirm an accidental cancellation between different three-loop terms, first observed in~\cite{Martin:2007pg}, and show that it persists to higher scales and becomes more effective with the inclusion of higher radiative corrections. Existing partial three-loop calculations that include only one of the two cancelling terms may overestimate the Higgs mass. We give analytic expressions for the three- and four-loop corrections in terms of Standard Model parameters and provide a complete dictionary for translating parameters between the SM and the MSSM and the $\MSbar$ and $\DRbar$ renormalization schemes.
\end{abstract}
%\vfil

%\end{titlepage}
\maketitle

%----------------------------------------------------------------------------------------
%	INTRODUCTION, OUTLINE OF PREVIOUS ANALYSES
%----------------------------------------------------------------------------------------

\section{Introduction} \label{sect:intro}

The discovery of the Higgs boson at the LHC by the ATLAS and CMS collaborations \cite{Chatrchyan:2012ufa, Aad:2012tfa} is a landmark achievement in high-energy physics. Combining the $h \rightarrow ZZ, \gamma\gamma$ decay channels, using  $\approx$ 5 fb${}^{-1}$ of data at $\sqrt{s} = 7$ TeV and $\approx$ 20 fb${}^{-1}$ of data at $\sqrt{s} = 8$ TeV, the Higgs boson mass is measured to be \cite{ATLAS-CONF-2013-014, CMS-PAS-HIG-13-005}
\ba{
\text{ATLAS: } & 125.5 \pm 0.2 \ {}^{+0.5}_{-0.6} \GeV , \\
\text{CMS: } & 125.7 \pm 0.3 \pm 0.3 \GeV ,
}
where the quoted uncertainties are statistical and systematic, respectively. 

It is by now well known that a variety of supersymmetric models can accommodate the observed Higgs mass and Standard Model (SM)-like couplings~\cite{Heinemeyer:2011aa}. One of the simplest possibilities for supersymmetry (SUSY) is that the Higgs boson is the lightest $CP$-even state $h$ in the minimal supersymmetric Standard Model (MSSM), and its mass, which is bounded at tree level by $m_Z$, receives large radiative corrections from heavy stop squarks. Exactly how heavy the stop squarks should be is a function of other model parameters, but if they are fixed, then the stop scale can be predicted. Since heavy-stop models are well motivated, it is of considerable interest to make the predictions precise, particularly in a handful of benchmark models. The stop mass scales in these benchmarks provide interesting targets for future experimental programs. 

Various methods have been employed to compute the Higgs mass to high precision in the MSSM. Broadly, the calculations fall into two categories: fixed-order computations in the full MSSM, and resummed (renormalization group or ``RG") analyses in effective theories.

Examples of fixed-order computations include the ``diagrammatic" method and the effective potential method. In the former, the renormalized self-energies appearing in the Higgs propagator matrix are evaluated from the complete set of Feynman diagrams up to a fixed-loop order~\cite{Haber:1990aw,Heinemeyer:1998jw, Martin:2004kr}. In the latter, radiative corrections to the Higgs masses are computed from derivatives of the MSSM potential $V(H_1, H_2)$ evaluated at the vacuum expectation values (vev) $\mean{H_1} = v_1, \, \mean{H_2} = v_2$~\cite{Haber:1993an, Espinosa:1999zm, Espinosa:2000df, Martin:2001vx, Martin:2002iu}. The effective potential result is obtained from the diagrammatic calculation in the zero external momentum approximation. Fixed-order computations have the virtue of being easily incorporated into numerical codes that accept arbitrary MSSM spectra, and have now been computed up to partial three-loop order~\cite{Martin:2007pg,Harlander:2008ju,Kant:2010tf,Feng:2013tvd}.

Effective field theory (EFT) analyses proceed by integrating out MSSM particles at their thresholds, running the effective theory couplings (most importantly the Higgs potential quartic couplings) down to the electroweak scale, and evaluating the Higgs pole mass or its effective potential approximation in the effective theory \cite{Casas:1994us, Carena:1995bx, Carena:1995wu, Haber:1996fp, Carena:2000dp}. This technique is most efficient in ``simplified models," where the MSSM decoupling can be performed at one or two scales, and below those scales the effective theory reduces to the SM. In the simplest case (``High-Scale SUSY"), the entire MSSM, including the second Higgs doublet, is decoupled simultaneously at a characteristic SUSY scale $M_S$. Calculations have been performed in this model using three-loop SM $\beta$ functions for the most important couplings~\cite{Degrassi:2012ry}. Furthermore, EFT methods may be used to obtain fixed-order formulas for the Higgs mass, by solving the RG equations analytically and perturbatively instead of numerically.

For low SUSY scales, where logarithmic radiative corrections are of size similar to the nonlogarithmic corrections, fixed-order computations are expected to be the most accurate, since they typically include a larger set of nonlogarithmic terms. For very high SUSY scales, the logs become large and fixed-order calculations break down, while EFT calculations remain trustworthy since they resum infinitely many large-log terms. For intermediate scales, where the logs are large enough to dominate but the perturbative series still exhibits converging behavior, one would expect both calculations to be valid, particularly if the fixed-order calculation is performed to high enough loop order.

One need only perform crude estimates to recognize that all three ranges of $M_S$ can be accessed by the benchmark heavy-stop models. Since most public codes utilize fixed order estimates for the Higgs mass, it is critical to understand the parameter regimes in which these estimates are trustworthy. The range of validity depends on the loop order, and for low orders can also depend strongly on the choice of renormalization scale.

In this paper, we compare fixed-order and resummed calculations in the cases of high-scale SUSY and a similar ``electrosplit" model where the Higgsinos and electroweak gauginos are allowed to be light\footnote{This is similar to split SUSY, but we keep the gluino as heavy as the scalars. We choose this somewhat unusual splitting in the gaugino sector for phenomenological rather than top-down reasons; the correction to the Higgs mass is largest for large $M_3$ and small $M_1$, $M_2$, $\mu$.}. By matching the MSSM onto the SM with two-loop threshold corrections and perturbatively solving the SM renormalization group equations (RGEs), we obtain three- and four-loop fixed-order formulas for $m_h$ that include terms through next-to-next-to-leading-log in the dominant couplings. We analyze the regimes of validity for these formulas and the impact of the higher-order corrections on the $m_h\rightarrow M_S$ prediction. We observe that convergence is better when couplings are evaluated at a renormalization scale equal to the SUSY scale rather than at the top quark mass, and that four-loop results fall within 0.5--1 GeV of the resummed calculation to scales of order a few tens of TeV.  Solving the RGEs numerically, for example at benchmark points with large $\tan\beta$ and small mixing in the stop sector, we find $M_S\approx 18\pm6$ TeV and $M_S\approx 7\pm2$ TeV for the heavy and light electroweakino cases, respectively. This result is in some tension with the results of~\cite{Feng:2013tvd}. The discrepancy may be due in part to a cancellation between three-loop terms at order $\alpha_s^2\alpha_t$ and $\alpha_t^2\alpha_s$, first noticed in~\cite{Martin:2007pg}; the $\alpha_t^2\alpha_s$ terms are absent from the calculation of~\cite{Feng:2013tvd}. We demonstrate that the cancellation persists at much higher SUSY scales than considered in~\cite{Martin:2007pg} and becomes even more effective with the inclusion of higher-order corrections. 

In addition to our quantitative results, we attempt to provide a contained dictionary for the translation of the parameters entering into the radiative corrections between different renormalization schemes and theories, so that our three- and four-loop NNLL formulas can be used in existing two-loop public codes. Although we consider models with only one or two decoupling scales, these capture the most significant higher-order corrections, and the formulas should give good approximations for more generic spectra.

This paper is organized as follows. In Sec. \ref{sect:matching}, we outline the matching procedure at the high scale $M_S$ and enumerate the threshold corrections to the running parameters. In Sec. \ref{sect:rge}, we give a brief overview of the renormalization group evolution in the SM and describe the perturbative solution that generates fixed-order analytic expressions for the radiative corrections to the Higgs mass. Readers interested primarily in final expressions can jump to Sec. \ref{sect:comparison}, where we present the fixed-order formulas for $m_h$. In this section we also compare the fixed-order estimates to the integration of the RGEs in benchmark models with small and large stop mixing and electroweakino masses.  We study the three- and four-loop contributions in detail. In Section \ref{sect:concl},  we conclude. Supporting technical details relevant to Sections II and III, including parameter conversion between the MSSM $\overline{\rm{DR}}$ and SM $\overline{\rm{MS}}$ schemes, are collected in appendices.

%----------------------------------------------------------------------------------------
%	INTEGRATING OUT THE HEAVY PARTICLES
%----------------------------------------------------------------------------------------

\section{Integrating Out the Heavy Particles} \label{sect:matching}

We begin with an overview of the threshold corrections to the running SM parameters in the $\MSbar$ scheme, obtained by integrating out the MSSM at a scale $M_S$. For the Higgs quartic coupling, we include one-loop gauge, Higgs, and third generation Yukawa corrections, as well as two-loop corrections controlled by the top Yukawa and strong gauge coupling. We pay particular attention to terms arising from changing the renormalization scheme from $\DRbar$ in the MSSM to $\MSbar$.

The quartic coupling in the MSSM is determined at leading order by the D-terms,
\be{
\l_{\text{tree}} = \frac14 (g_2^2 + g_Y^2) c_{2\b}^2 , \label{eqn:lambdatree}
} 
where, in this section, we use the notation $\l \equiv \l_{\text{MSSM}} (M_S)$ for the MSSM quartic coupling in the $\MSbar$ scheme at $Q = M_S$ and $\cb = \cos\beta, \sb = \sin\beta$, and $\tb = \tan\beta = v_u/v_d$, with $v_u$ and $v_d$ the vacuum expectation values of the MSSM Higgs doublets. It is well known that $\l$ receives significant nonlogarithmic radiative corrections from the mixing of heavy SUSY partners at the high scale. In the framework of effective field theory, these ``threshold corrections'' are a result of the decoupling of heavy particles at the high scale. 

The largest effect comes from the top-stop sector. The squark mass matrix in the MSSM has the form
\be{
\mathcal{M}_{\tt}^2 = \Bigg( \begin{array}{cc} 
m_{\tt_L}^2 + m_t^2 + c_{2\b} \Big(\frac12 - \frac23 s_W^2 \Big) m_Z^2 & m_t X_t \\ 
m_t X_t & m_{\tt_R}^2 + m_t^2 + \frac23 c_{2\b} s_W^2 m_Z^2
\end{array} \Bigg) ,
}
where we have followed the notation of \cite{Carena:2000dp} with the stop mixing parameter defined as $X_t = A_t - \mu \cot\beta$ and $s_W = \sin\q_W$, with $\q_W$ the Weinberg angle. We will set all $CP$-violating phases in the MSSM to zero. Diagonalizing this matrix yields the tree-level stop masses $m_{\tt_1}, m_{\tt_2}$ and the stop mixing angle $\q_{\tt}$. Neglecting the terms proportional to $m_Z$ and setting $m_{\tt_L} = m_{\tt_R} = \MSUSY, M_S^2 = \MSUSY^2 + m_t^2$, we obtain the simplified squark mass matrix
\be{
\mathcal{M}_{\tt}^2 = \Bigg( \begin{array}{cc} 
M_S^2 & m_t X_t \\ 
m_t X_t & M_S^2
\end{array} \Bigg) ,
} 
with
\begin{alignat}{2}
m_{\tt_{1,2}}^2 &= M_S^2 \mp |m_t X_t| .%, & \\
%\q_{\tt} &= \begin{cases} \pi/4 & X_t > 0 \\ -\pi/4 & X_t < 0 \end{cases} .
\end{alignat}
We choose the scale $M_S$ as our high scale, assuming that all supersymmetric partners have similar masses; however, we keep the MSSM $\mu$ parameter free with $\mu=M_1=M_2$ so that light electroweakinos can be accommodated.

From \cite{Carena:1995bx, Carena:2000dp}, we include the most relevant one-loop corrections that include terms from decoupling stops, sbottoms, and staus:
\ba{
\D^{(\a_t)}_{\text{th}} \l &= 6\k h_t^4 \sb^4 \XtMS^2 \Big(1 - \frac{\XtMS^2}{12} \Big) + \frac34 \k h_t^2 \sb^2 (g_2^2 + g_Y^2) \XtMS^2 c_{2\b} , \label{eqn:1loopstop} \\
\D^{(\a_b)}_{\text{th}} \l &= - \frac12 \k h_b^4 \sb^4 \muMS^4 , \label{eqn:1loopsbot} \\
\D^{(\a_\tau)}_{\text{th}} \l &= - \frac16 \k h_\tau^4 \sb^4 \muMS^4 , \label{eqn:1loopstau}
} 
where $h_t$ ($h_b, h_\tau$) is the MSSM top (bottom, tau) Yukawa coupling, $\XtMS = X_t/M_S$, $\muMS = \mu/M_S$, and following the notation of \cite{Martin:2007pg}, we keep track of loop order via $\k = 1/(16\pi^2)$. Note that the parameters on the right-hand sides of these equations are $\MSbar$ running couplings evaluated at $M_S$. At tree level, the MSSM Yukawa couplings are related to the SM Yukawa couplings by
\be{
y_t = h_t \sb , \qquad y_b = h_b \cb, \qquad y_\tau = h_\tau \cb ;
}
however, these couplings are modified at one-loop order at $M_S$ by \cite{Guasch:2001wv, Carena:2000yi}:
\ba{
h_t &= \frac{y_t}{\sb} \frac1{1 - \k (\D h_t + \cot\b \, \d h_t)} , \label{eqn:hytMSbar} \\
h_b &= \frac{y_b}{\cb} \frac1{1 - \k (\D h_b + \tb \, \d h_b)} , \label{eqn:hybMSbar}  \\
h_\tau &= \frac{y_\tau}{\cb} \frac1{1 - \k \tb \, \d h_\tau} , \label{eqn:hytauMSbar}
}
where
\ba{
\D h_t &= \frac83 g_3^2 m_{\tilde{g}} X_t \, I(m_{\tt_1}, m_{\tt_2}, m_{\tilde{g}})
- h_b^2 \mu \cot\b X_b \, I(m_{\tilde{b}_1}, m_{\tilde{b}_2}, \mu) , \label{eqn:Dht} \\
\d h_t &=  g_2^2 M_2 \mu \Big( [c_b^2 I(m_{\tilde{b}_1},  M_2, \mu) + s_b^2 I(m_{\tilde{b}_2},  M_2, \mu)]
+ \frac12 [c_t^2 I(m_{\tt_1}, M_2, \mu) +  s_t^2 I(m_{\tt_2}, M_2, \mu)] \Big) \nl
& \quad + \frac13 g_Y^2 M_1 \Big( \frac23 X_t \tb I( m_{\tt_1}, m_{\tt_2}, M_1) 
- \frac12 \mu [c_t^2 I(m_{\tt_1}, M_1, \mu) + s_t^2 I(m_{\tt_2}, M_1, \mu)] \nl
& \quad + 2\mu [s_t^2 I(m_{\tt_1}, M_1, \mu) + c_t^2 I(m_{\tt_2}, M_1, \mu)] \Big) , \label{eqn:dEWht}
}
\ba{
\D h_b &= \frac83 g_3^2 m_{\tilde{g}} X_b \, I(m_{\tilde{b}_1}, m_{\tilde{b}_2}, m_{\tilde{g}})
- h_t^2 \mu \tb X_t \, I(m_{\tt_1}, m_{\tt_2},  \mu ) , \label{eqn:Dhb} \\
\d h_b &= g_2^2 M_2 \mu \Big( [c_t^2 I(m_{\tt_1}, M_2, \mu) +  s_t^2 I(m_{\tt_2}, M_2, \mu)]
+ \frac12 [c_b^2 I(m_{\tilde{b}_1},  M_2, \mu) + s_b^2 I(m_{\tilde{b}_2},  M_2, \mu)] \Big) \nl
& \quad + \frac13 g_Y^2 M_1 \Big( -\frac13 X_b \cot\b \, I( m_{\tilde{b}_1}, m_{\tilde{b}_2}, M_1) 
+ \frac12 \mu [c_b^2 I(m_{\tilde{b}_1}, M_1, \mu) + s_b^2 I(m_{\tilde{b}_2}, M_1, \mu)] \nl
& \quad + \mu [s_b^2 I(m_{\tilde{b}_1}, M_1, \mu) + c_b^2 I(m_{\tilde{b}_2}, M_1, \mu)] \Big) , \label{eqn:dEWhb}
}
\ba{
\d h_\tau &= g_2^2 M_2 \mu \Big( I(m_{\tilde{\nu}_\tau}, M_2, \mu) 
+ \frac12 [c_\tau^2 I(m_{\tilde{\tau}_1},  M_2, \mu) + s_\tau^2 I(m_{\tilde{\tau}_2},  M_2, \mu)] \Big) \nl
& \quad - g_Y^2 M_1 \Big( X_\tau \cot\b \, I( m_{\tilde{\tau}_1}, m_{\tilde{\tau}_2}, M_1) 
+ \frac12 \mu [c_\tau^2 I(m_{\tilde{\tau}_1}, M_1, \mu) + s_\tau^2 I(m_{\tilde{\tau}_2}, M_1, \mu)] \nl
& \quad - \mu [s_\tau^2 I(m_{\tilde{\tau}_1}, M_1, \mu) + c_\tau^2 I(m_{\tilde{\tau}_2}, M_1, \mu)] \Big) . \label{eqn:dEWhtau}
}
%\ba{
%h_t = \frac{y_t}{\sb} \Bigg\{ 1 &
%- \frac83 \k g_3^2 \XtMS \, I\Big(1 - \frac{m_t}{M_S} \XtMS, 1 + \frac{m_t}{M_S} \XtMS, 1\Big)
%+ \k h_b^2 \frac{\muMS}{\tb} \wh{X}_b \, I(1, 1, \mu^2) 
%\Bigg\}^{-1} \\
%h_b = \frac{y_b}{\cb} \Bigg\{ 1 &
%- \frac43 \k g_3^2 \wh{X}_b 
%+ \k h_t^2 \muMS \tb \XtMS \, I\Big(1 - \frac{m_t}{M_S} \XtMS, 1 + \frac{m_t}{M_S} \XtMS, \muMS^2\Big) \nl
%& - \frac12 \k g_2^2 \muMS \tb \Big[ I\Big(1 - \frac{m_t}{M_S} \XtMS, 1, \muMS^2\Big) + I\Big(1 + \frac{m_t}{M_S} \XtMS, 1, \muMS^2\Big) + I(1, 1, \muMS^2) \Big] \nl
%& - \frac13 g_1^2 \muMS \tb \Big[ \frac16 + \frac32 I(1, 1, \muMS^2) \Big]
%\Bigg\}^{-1} , \\
%h_\tau = \frac{y_\tau}{\cb} \Bigg\{ 1 &
%-\frac32 \k g_2^2 \muMS \tb \, I(1, 1, \muMS^2) + \frac12 \k g_1^2 \muMS \tb \Big[ 1 - I(1, 1, \muMS^2) \Big]
%\Bigg\}^{-1}, 
%}
Here $X_b = A_b  - \mu \, \tb$ and $X_\tau = A_\tau  - \mu \, \tb$ are the sbottom and stau mixing parameters, $s_t, s_b, s_\tau$ $(c_t, c_b, c_\tau)$ are the sines (cosines) of the stop, sbottom, and stau mixing angles, and the function $I(a,b,c)$ is defined as
\be{
I(a,b,c) = \frac{a^2 b^2 \log(a^2/b^2) + b^2 c^2 \log(b^2/c^2) + a^2 c^2 \log(c^2/a^2)}{(a^2 - b^2)(b^2 - c^2)(a^2 - c^2)} .
}
We will set all MSSM masses $m_{\tilde{g}} = m_{\tilde{b}_i} = m_{\tilde{\tau}_i} = m_{\tilde{\nu}_i} = M_S$ (such that $s_X^2 = c_X^2 = 1/2$ with $X = t, b, \tau$), assume $A_t = A_b = A_\tau$, and consider the two scenarios $M_2 = M_1 = \mu = M_S$ (the ``high $\mu$" case) and $M_2 = M_1 = \mu = 200$ GeV (the ``low $\mu$" case).\footnote{We have neglected the threshold corrections from this intermediate scale to $\l, y_t$. They can be found in \cite{Giudice:2011cg}, and involve only $g_Y, g_2, \l$. We estimate that the corrections to $\l$ lower $m_h$ by about $0.5$ GeV.} Taking the appropriate limits when the arguments are degenerate, we have the common asymptotic forms for $I(a,b,c)$:
\begin{alignat}{2}
I(M_S, M_S, M_S) &= \frac1{2M_S^2} , &&  \\
I(M_S, M_S, \mu) &= \frac1{M_S^2} \frac1{1-\muMS^2} \Big[1 + \frac{\muMS^2 \log\muMS^2}{1 - \muMS^2} \Big] ,  && \qquad \muMS < 1 , \\
I(M_S, \mu, \mu) &= -\frac1{M_S^2} \frac1{1-\muMS^2} \Big[1 + \frac{\log\muMS^2}{1 - \muMS^2} \Big] , && \qquad \muMS < 1 . 
\end{alignat}

The expressions for the dominant two-loop corrections of $\mathcal{O}(\a_s \a_t)$ and $\mathcal{O}(\a_t^2)$ will depend on the scheme used for the one-loop corrections. The two-loop finite $\mathcal{O}(\a_s \a_t)$ corrections were computed diagrammatically in the OS scheme in \cite{Heinemeyer:1998jw}, and in the $\DRbar$ scheme using the effective potential method in \cite{Espinosa:1999zm}. In a followup to the latter paper \cite{Espinosa:2000df}, the $\mathcal{O}(\a_t^2)$ corrections were also computed. It was shown in \cite{Espinosa:2000df} and \cite{Carena:2000dp} that the different expressions for the $\mathcal{O}(\a_t \a_s)$ corrections in the two schemes are reconciled once the one-loop $\mathcal{O}(\a_t)$ corrections are written in the appropriate scheme.

We will express $\l$ in terms of the MSSM couplings in the $\MSbar$ scheme given in Eqs.~(\ref{eqn:hytMSbar}), (\ref{eqn:hybMSbar}), (\ref{eqn:hytauMSbar}). To determine $\D^{(\a_s \a_t)}_{\text{th}} \l, \D^{(\a_t^2)}_{\text{th}} \l$ in this scheme, let us write the one-loop correction to the running $\DRbar$ Higgs mass obtained from the Higgs effective potential in \cite{Espinosa:2000df}
\be{
\D^{(\a_t)}_{\DRbar} m_h^2 = \frac{3}{2\pi^2} \frac{\wt{m}_t^4}{\tilde{v}^2} \Bigg\{ \log\Big(\frac{\wt{M}_S^2}{\wt{m}_t^2}\Big) + \frac{\wt{X}_t^2}{\wt{M}_S^2} \Big(1 - \frac1{12} \frac{\wt{X}_t^2}{\wt{M}_S^2} \Big) \Bigg\} , \label{eqn:mh21loopDR}
}
where we have used the notation of Table~\ref{table:schemenotation} in Appendix~\ref{app:schemes}, i.e. all parameters with a tilde are in the $\DRbar$ scheme and evaluated at a renormalization scale $Q$. Here, we have included the logarithmic contribution; in the effective theory, this is obtained from the running below $M_S$.
Parameters in the logarithmic term should be converted to the $\MSbar$ scheme in the SM, i.e. $\wt{m}_t (M_S) \rightarrow \mb_t (M_S)$, multiplied by the appropriate one-loop corrections given in Appendix \ref{app:schemes}. This substitution produces a finite two-loop correction once the logarithm is expanded to one-loop order. For the nonlogarithmic terms, we change $\wt{m}_t \rightarrow m_t, \wt{X}_t \rightarrow X_t, \wt{M}_S \rightarrow M_S$, all at $Q = M_S$, to match the threshold corrections in Eq. (\ref{eqn:1loopstop}). After performing the scheme conversion for the one-loop terms and modifying the two-loop $\mathcal{O}(\a_s \a_t)$ and $\mathcal{O}(\a_t^2)$ terms in \cite{Espinosa:2000df}, we find for the threshold corrections to $\l$:
\ba{
\D^{(\a_s \a_t)}_{\text{th}} \l = 16\k^2 h_t^4 \sb^4 g_3^2 \Bigg\{ &
-2\XtMS + \frac13 \XtMS^3 - \frac1{12} \XtMS^4 
\Bigg\}, \label{eqn:2loopasat}
}
\ba{
\D^{(\a_t^2)}_{\text{th}} \l = 3\k^2 h_t^6 \sb^4 \Bigg\{ &
- \frac32 + 6\muMS^2 - 2(4 + \muMS^2) f_1(\muMS) +3\muMS^2 f_2(\muMS) + 4 f_3(\muMS) - \frac{\pi^2}3 \nl
& + \Big[ -\frac{17}2 - 6\muMS^2 - (4 + 3\muMS^2) f_2(\muMS) + (4 - 6\muMS^2) f_1(\muMS) \Big] \XtMS^2 \nl
& + \Big[ 23 + 4\sb^2 + 4\muMS^2 + 2 f_2(\muMS) - 2(1 - 2\muMS^2) f_1(\muMS) \Big] \frac{\XtMS^4}4 
- \frac{13}{24} \XtMS^6 \sb^2 \nl
& + \cb^2 \Bigg[ - \frac92 + 60K + \frac{4\pi^2}3 + \Big( \frac{27}2 - 24k \Big)\XtMS^2 - 6\XtMS^4 \nl
& \qquad \quad - (3 + 16K)(4\XtMS + \wh{Y}_t) \wh{Y}_t + 4(1 + 4K)\XtMS^3 \wh{Y}_t \nl
& \qquad \quad + \Big( \frac{14}3 + 24K) \XtMS^2 \wh{Y}_t^2 - \Big( \frac{19}{12} + 8K) \XtMS^4 \wh{Y}_t^2 \Bigg] 
\Bigg\} . \label{eqn:2loopat2}
}
We have borrowed the notation of \cite{Espinosa:2000df}, with the constant $K$, parameter $\wh{Y}_t$, and functions $f_i$ defined as
\ba{
K &= - \frac1{\sqrt{3}} \int_0^{\pi/6} dx \log(2\cos x) \sim -0.1953256 , \\
\wh{Y}_t &= (A_t - \mu \tb)/M_S = \XtMS + \frac{2\muMS}{\sin2\b}, \\
f_1(\muMS) &= \frac{\muMS^2}{1 - \muMS^2} \log\muMS^2 , \\
f_2(\muMS) &= \frac1{1 - \muMS^2} \Big[ 1 + \frac{\muMS^2}{1 - \muMS^2} \log\muMS^2 \Big] , \\
f_3(\muMS) &= \frac{-1 + 2\muMS^2 + 2\muMS^4}{(1 - \muMS^2)^2} \Bigg[ \log\muMS^2 \log(1 - \muMS^2) + Li_2 (\muMS^2) - \frac{\pi^2}6 - \muMS^2 \log\muMS^2 \Bigg] ,
}
and the dilogarithm function $Li_2$ is 
\be{
Li_2 (x) = - \int_0^1 dy \frac{\log(1 - xy)}y .
}
We will be interested in the limits of the $f_i$ as $\muMS \rightarrow$ 0 or 1, with
\be{
f_{(1,2,3)}(\muMS) = \left\{ 
\begin{array}{cc} 
(0, 1, \frac{\pi^2}6) \quad & \quad \muMS = 0 , \\
(-1, \frac12, -\frac94) \quad & \quad \muMS = 1 . 
\end{array} \right.
}
Finally, we include one-loop threshold corrections from converting the tree-level quartic coupling from the $\DRbar$ to the $\MSbar$ scheme and those from the heavy Higgs bosons, which are taken from \cite{Giudice:2011cg}:
\ba{
\D_{\text{th}}^{(\text{sc})} \l &= - \k \Big[ \Big( \frac34 - \frac16 c_{2\b}^2 \Big) g_2^4 + \frac1{2} g_Y^2 g_2^2 + \frac14 g_Y^4 \Big] , \label{eqn:1loopsch} \\
\D_{\text{th}}^{(H)} \l &= - \frac1{16} \k (g_2^2 + g_Y^2)^2 s_{4\b}^2 . \label{eqn:1loopH}
}
Our final expression for $\l_{\text{MSSM}} (M_S)$ to which we match the SM running quartic coupling is
\ba{
\l_{\text{MSSM}} (M_S) = \l_{\text{tree}} & + \D_{\text{th}}^{(\text{sc})} \l + \D_{\text{th}}^{(H)} \l + \D^{(\a_t)}_{\text{th}} \l + \D^{(\a_b)}_{\text{th}} \l + \D^{(\a_\tau)}_{\text{th}} \l \nl
& + \D^{(\a_s \a_t)}_{\text{th}} \l  + \D^{(\a_t^2)}_{\text{th}} \l . \label{eqn:lambdaMSSM}
}

%----------------------------------------------------------------------------------------
%	RENORMALIZATION GROUP EVOLUTION
%----------------------------------------------------------------------------------------

\section{Running the SM Down from $M_S$} \label{sect:rge}

Once the heavy sparticles have been integrated out, the SM parameters can be run down to the electroweak scale and the spectrum computed. The $\beta$-function $\b_\l = \dt{\l}$ for a generic running coupling $\l$ can be written as
\be{
\b_\l (t) % \sum_{n=1}^{\infty} \k^n \b_\l^{(n)} (t) 
= \sum_{n=1}^\infty \k^n \sum_{k=0}^\infty \frac{\b_\l^{(n,k)}(\tt)}{k!} (t - \tt)^k , \label{eqn:betaexpansion}
}
where
\be{
\k \equiv \frac1{16\pi^2} , \qquad 
t \equiv \log Q , \qquad 
\b_\l^{(n,k)} (t) \equiv \frac{\mathrm{d}^k \b_\l^{(n)}}{\mathrm{d} t^k} (t) .
}
We will also use the shorthand $\b_\l^{(n)} \equiv \b_\l^{(n,0)}$. We will denote $\wt{Q}$ as the high scale, and we define $L \equiv \tt - t = \log(\wt{Q}/Q)  > 0$. Integrating from $t$ to $\tt$, we find
\ba{
\l(Q) = \l(\wt{Q}) - \sum_{n=1}^{\infty} \k^n \sum_{k=0}^\infty (-1)^k \frac{\b_\l^{(n,k)} (\tt)}{(k+1)!} \, L^{k+1} .
\label{eqn:lambdaQhigh}
}
Alternatively, we can expand the beta-function coefficients $\b_\l^{(n,k)}$ about the low scale $Q$,
\ba{
\l(\wt{Q}) = \l(Q) + \sum_{n=1}^{\infty} \k^n \sum_{k=0}^\infty \frac{ \b_\l^{(n,k)} (t)}{(k+1)!} \, L^{k+1} . 
\label{eqn:lambdaQlow}
}
To see the equivalence with Eq.~(\ref{eqn:lambdaQhigh}), we can evolve the beta-function coefficients $\b_\l^{(n,k)} (\tt)$  down to the low scale $\b_\l^{(n,k)} (t)$ using the same expansion as in Eq.~(\ref{eqn:betaexpansion}). The effect on the beta-functions in Eq.~(\ref{eqn:lambdaQhigh}) is to remove the tildes and make all the leading signs negative, which agrees with Eq.~(\ref{eqn:lambdaQlow}).

% TABLE: RGE ORDERS
\begin{table}[htb]
\begin{center}
\begin{tabular}{ccc}
Parameter \qquad & \quad $\b$-function order, resummation \quad & \quad $\b$-function order, fixed-order \quad \\ 
\hline
$g_3$ & 3 + 4-loop QCD & (2, 0) \\
$y_t$ & 3 & (3, 1) \\
$\l$ & 3 & (3, 1) \\
$g_1, g_2$ & 3 & 1 in Eq.~(\ref{eqn:lambdatree}) \\
$y_b, y_\tau$ & 2 & $y_b, y_\tau$: (1,1) in Eqs.~(\ref{eqn:hybMSbar}-\ref{eqn:hytauMSbar})
\end{tabular}
\end{center}
\caption{Orders of the $\b$-functions of SM parameters used in solving the RGEs in the resummation and fixed-order methods. The second digit of the 2-tuple in the fixed-order column indicates at which order electroweak, bottom, and tau contributions are included. See section \ref{sect:comparison} for more details on the fixed-order calculation. The $\beta$-functions are taken from \cite{Buttazzo:2013uya}, and we have checked them to 2-loop order against \cite{Arason:1991ic} with corrections in \cite{Luo:2002ey}.}
\label{table:RGEorders}
\end{table}

We use two different methods to perform the renormalization group running. The most precise approach is to numerically integrate the coupled SM $\MSbar$ RGEs between $Q = M_t$ and $Q = M_S$ for the seven parameters ${g_3, g_2, g_1, y_t, y_b, y_\tau, \l}$, with $g_1 = \sqrt{5/3} g_Y$ the SM hypercharge coupling expressed in the $SU(5)$ normalization. In the middle column of Table~\ref{table:RGEorders} we indicate the order of $\beta$-function used for each coupling.
Observables and electroweak scale boundary values for the SM parameters are taken from Tables~2 and~3 of \cite{Buttazzo:2013uya}. We reproduce the observables and the parameters $g_2, g_1, y_b,$ and $y_\tau$ in Tables~\ref{table:SMobs}~and~\ref{table:SMparams}. The next-to-next-to-leading-order (NNLO) values of $g_3$ and $y_t$ are given in terms of the observables $M_t$ and $\a_s(M_Z)$ in \cite{Buttazzo:2013uya}, to which we refer the reader for further details:
\ba{
y_t(Q = M_t) &= 0.93697 \pm 0.00550 \Big( \frac{M_t}{\GeV} - 173.35 \Big) - 0.00042 \frac{\a_s(M_Z) - 0.1184}{0.0007} , \label{eqn:ytNNLO} \\
g_3(Q = M_t) &= 1.1666 + 0.00314 \frac{\a_s(M_Z) - 0.1184}{0.0007} - 0.00046 \Big( \frac{M_t}{\GeV} - 173.35 \Big) \label{eqn:g3NNLO} .
}
We note that the central value for $y_t(M_t)$ quoted here includes the N${}^3$LO pure QCD contribution. The value of $\l(M_t)$ is determined by beginning with the approximate value of $\l(M_t)$ corresponding to the Higgs pole mass $M_h \sim 125.6 \GeV$. The numerical integration yields a value $\bar{\l} (M_S)$. This is compared to Eq. (\ref{eqn:lambdaMSSM}) from Section \ref{sect:matching}, which is determined by the other couplings at $M_S$. If the difference exceeds a specified tolerance, the starting value $\l(M_t)$ is appropriately adjusted. This procedure is iterated until convergence is achieved. We find that for a tolerance of $10^{-6}$, about 10 iterations are required.

% TABLE: SM OBSERVABLES
\begin{table}[htb]
\begin{center}
\begin{tabular}{ll}
Observable \qquad & \qquad Value \qquad \\ 
\hline
$SU(3)_c$ $\MSbar$ gauge coupling (5 flavors) & $\a_s(M_Z) = 0.1184 \pm 0.0007$ \\
Fermi constant from muon decay & $V = (\sqrt2 G_F)^{-1/2} = 246.21971 \pm 0.00006 \GeV$ \\
Top quark pole mass & $M_t = 173.36 \pm 0.65 \pm 0.3 \GeV$ \\
Z boson pole mass & $M_Z = 91.1876 \pm 0.0021 \GeV$ \\
Higgs pole mass & $M_h = 125.66 \pm 0.34 \GeV$
\end{tabular}
\end{center}
\caption{SM observables, collected in Table 2 of \cite{Buttazzo:2013uya}.}
\label{table:SMobs}
\end{table}

The second method is to solve the RGEs perturbatively around a reference scale. The result is a fixed-order expression. We take two values for the renormalization scale in this approach, $Q = M_S$ and $Q = M_t$. Since we know $\b_\l$ up to the three-loop level, we can write an expansion up to four-loop order excluding only the four-loop N${}^3$LL terms, which we expect are small for large $M_S$:
\ba{
\l(M_t) = \l(M_S) &- \k \b_\l^{(1)} (M_S) \, L 
- \k^2 \b_\l^{(2)} (M_S) \, L + \k \frac{\b_\l^{(1,1)} (M_S)}{2!} \, L^2 \nl
& - \k^3 \b_\l^{(3)} (M_S) \, L + \k^2 \frac{\b_\l^{(2,1)} (M_S)}{2!} L^2 - \k \frac{\b_\l^{(1,2)} (M_S)}{3!} \, L^3 \nl
& + \k^3 \b_\l^{(3,1)} (M_S) \, L^2 - \k^2 \frac{\b_\l^{(2,2)} (M_S)}{3!} L^3 + \k \frac{\b_\l^{(1,3)} (M_S)}{4!} \, L^4 + \ldots. \label{eqn:lambdaQMS}
}
Note that the derivatives $\b_\l^{(n,k)}, k > 0$, contain $\b$-functions for the couplings that appear in $\b_\l^{(n)}$. The computations for both choices of renormalization scale are truncated at four-loop order; however, for $Q = M_S$, the truncation occurs before the couplings $y_t (M_S)$ $[g_3 (M_S)]$ are computed,
 %using a three-loop [two-loop] fixed-order calculation
and vice versa for $Q = M_t$. The results for the two choices should converge with the addition of higher-order $\b_\l^{(n)}$.

Appendix \ref{app:betafuncs} contains the relevant $\b$-functions appearing in Eq. (\ref{eqn:lambdaQMS}). We have included the $g_1, g_2, y_b, y_\tau$ terms in $\b_\l^{(1)}, \b_{y_t}^{(1)}$. For larger $M_S$, the electroweak terms grow in importance since the values of $g_1, g_2$ change much more slowly compared to $y_t$. Their inclusion in $\b_{y_t}^{(1)}$ lowers $M_h$ by about 1 GeV, as the dominant term in $\b_{\l}^{(1)}$ is proportional to $y_t^4$. 

% TABLE: SM PARAMETERS AT Q = M_t
\begin{table}[htb]
\begin{center}
\begin{tabular}{ccc}
Parameter \qquad & \quad Value \quad \\ 
\hline
$g_2$ & 0.6483 \\
$g_Y = \sqrt{3/5} g_1$ & 0.3587 \\
$y_b$ & 0.0156 \\
$y_\tau$ & 0.0100 
\end{tabular}
\end{center}
\caption{Values of SM parameters at $Q = M_t$ using two-loop (NNLO) renormalization group running in the $\MSbar$ scheme, from Table 3 of \cite{Buttazzo:2013uya}. The $SU(5)$ normalization relates $g_1$ to the SM hypercharge coupling $g_Y$. We have used the two-loop 5-flavor $\MSbar$ renormalization group equations in the broken phase from \cite{Arason:1991ic} to run $m_b, m_\tau$ from their initial values $m_b (m_b) = 4.18 \GeV, M_\tau = 1.777 \GeV$ \cite{PDG}.}
\label{table:SMparams}
\end{table}

%----------------------------------------------------------------------------------------
%	FIXED-ORDER RESULT AND COMPARISON TO RESUMMATION
%----------------------------------------------------------------------------------------

\section{Fixed-Order Result and Comparison to Resummation} \label{sect:comparison}

In this section we present approximate three and four-loop NNLL fixed-order formulas for $m_h$ and compare to the result of numerical resummation.

The running Higgs mass at $M_t$ is given by
\be{
m_h^2 (M_t) = \l (M_t) v^2 (M_t) . \label{eqn:mhrun}
}
We use one-loop running to obtain $v(M_t) = 246.517$ GeV from $v(M_Z) \sim V$ (see Table~\ref{table:SMobs}). The logarithmic factors are $L = \log(M_S/M_t)$ and $L_\mu = \log(M_S/\mu)$ (note that the latter also includes logs of the form $\log(M_S/M_{1,2})$). Below, all parameters are in the $\MSbar$ scheme and should be evaluated at $Q = M_S$: 
\be{
\l(M_t) = \l + \k \d_1 \l + \k^2 \d_2 \l + \k^3 \d_3 \l + \k^4 \d_4 \l , \label{eqn:lambdaanalytic}
}
and
\ba{
\d_1 \l = \Bigg\{ & -12 \lambda ^2 - \lambda \Big[ 12 y_t^2 + 12 y_b^2 + 4 y_\tau^2 - 9 g_2^2 - \frac95 g_1^2 \Big] + 12 y_t^4 + 12 y_b^4 + 4 y_{\tau }^4  \nl
& \quad - \frac94 g_2^4 - \frac{9}{10} g_2^2 g_1^2 - \frac{27}{100} g_1^4 \Bigg\} L \nl
+ \ \Bigg\{ & - 6\l \Big[ g_2^2 + \frac15 g_1^2 \Big] + \Big[ g_2^2 + \frac35 g_1^2 \Big]^2 + 4g_2^4 \Big[ 1 - 2 \sb^2 \cb^2 \Big]  \Bigg\} L_\m, \label{eqn:d1lambda} \\
\d_2 \l = \Bigg\{ & 144 \lambda ^3 + \lambda^2 \Big[ 216 y_t^2 %+ 144 y_b^2 + 48 y_\tau^2 
- 108g_2^2 - \frac{108}5 g_1^2 \Big] % \nl
+ \l \Big[ -18 y_t^4 %- 144 y_b^4 - 48 y_\tau^4
+ 27 g_2^4 + \frac{54}5 g_2^2 g_1^2 + \frac{81}{25} g_1^4 \Big] \nl 
& + \l y_t^2 \Big[ %90 y_b^2 + 36 y_\tau^2 
- 96 g_3^2 - 81 g_2^2 - 21 g_1^2 \Big] %\nl
+ y_t^4 \Big[ -180 y_t^2 %- 36 y_b^2 - 24 y_\tau^2 
+ 192 g_3^2 + 54 g_2^2 + \frac{102}5 g_1^2 \Big] \nl
& + y_t^2 \Big[ %-72 y_b^4 - 24 y_\tau^4 +
\frac{27}2 g_2^4 + \frac{27}5 g_2^2 g_1^2 + \frac{81}{50} g_1^4 \Big] \Bigg\} L^2 \nl
- \ \Bigg\{ & \Bigg[24 \lambda + 12 y_t^2- 9 g_2^2 - \frac95 g_1^2 \Bigg] 
\Bigg[ 6 \l \Big[ g_2^2 + \frac15 g_1^2 \Big]^2 - \Big[ g_2^2 + \frac35 g_1^2 \Big]^2 - 4g_2^4 \Big[ 1 - 2 \sb^2 \cb^2 \Big] \Bigg] \Bigg\} L L_\m \nl
+ \Bigg\{ & 3 \Big[ g_2^2 + \frac15 g_1^2 \Big]  \Bigg[ 6 \l \Big[ g_2^2 + \frac15 g_1^2 \Big]^2 - \Big[ g_2^2 + \frac35 g_1^2 \Big]^2 - 4g_2^4 \Big[ 1 - 2 \sb^2 \cb^2 \Big] \Bigg] \Bigg\} L_\m^2 \nl
+ \ \Bigg\{ & 78 \lambda ^3 + 72 \lambda ^2 y_t^2 + \lambda y_t^2 ( 3 y_t^2 - 80 g_3^2) - 60 y_t^6 + 64 g_3^2 y_t^4  \Bigg\} L  , \label{eqn:d2lambda}
}
\ba{
\d_3 \l = \Bigg\{ & -1728 \lambda^4 - 3456 \lambda^3 y_t^2 + \lambda^2 y_t^2 (-576 y_t^2 + 1536 g_3^2) \nl
& + \lambda y_t^2 (1908 y_t^4 + 480 y_t^2 g_3^2 - 960 g_3^4) + y_t^4 (1548 y_t^4 - 4416 y_t^2 g_3^2 + 2944 g_3^4) \Bigg\} L^3 \nl
+ \ \Bigg\{ & -2340 \lambda^4 -3582 \lambda^3 y_t^2 + \lambda ^2 y_t^2 (-378 y_t^2 + 2016 g_3^2) \nl
& + \lambda y_t^2 (1521 y_t^4 + 1032 y_t^2 g_3^2 - 2496 g_3^4) + y_t^4 (1476 y_t^4 - 3744 y_t^2 g_3^2 + 4064 g_3^4) \Bigg\} L^2 \nl
+ \ \Bigg\{ & -1502.84 \lambda^4-436.5 \lambda ^3 y_t^2 - \lambda^2 y_t^2 (1768.26 y_t^2 + 160.77 g_3^2) \nl
& + \lambda y_t^2 (446.764 \lambda  y_t^4 + 1325.73 y_t^2 g_3^2 - 713.936 g_3^4) \nl
& + y_t^4 (972.596 y_t^4 - 1001.98 y_t^2 g_3^2 + 200.804 g_3^4) \Bigg\} L , \label{eqn:d3lambda}
}
\ba{
\d_4 \l = \Bigg\{ & 20736 \lambda^5 + 51840 \lambda^4 y_t^2 + \lambda^3 y_t^2 (21600 y_t^2 - 23040 g_3^2) \nl
& + \lambda^2 y_t^2 (-30780 y_t^4 - 18720 g_3^2 y_t^2 + 14400 g_3^4) \nl
& + \lambda y_t^2 (-22059 y_t^6 + 28512 g_3^2 y_t^4 +10560 g_3^4 y_t^2 - 10560 g_3^6) \nl
& + y_t^4 (-8208 y_t^6 + 56016 y_t^6 g_3^2 - 84576 y_t^2 g_3^4 + 44160 g_3^6) \Bigg\} L^4 \nl
+ \ \Bigg\{ & 48672 \lambda^5 + 101808 \lambda^4 y_t^2 + \lambda^3 y_t^2 (30546 y_t^2 - 49152 g_3^2 y_t^2) \nl
& \lambda^2 y_t^2 (-50292 y_t^4 - 40896 y_t^2 g_3^2 + 45696 g_3^4) \nl
& + \lambda y_t^2 (-33903 y_t^6 + 41376 y_t^4 g_3^2 + 35440 g_3^4 y_t^2 - 45184 g_3^6) \nl
& + y_t^4(-15588 y_t^6 + 86880 y_t^4 g_3^2 - 161632  y_t^2 g_3^4 + 112256 g_3^6) \Bigg\} L^3 \nl
+ \ \Bigg\{ & 63228.2 \lambda^5 +72058.1 \lambda^4 y_t^2 + \lambda ^3 y_t^2 (25004.6 y_t^2 - 11993.5 g_3^2) \nl
& + \lambda^2 y_t^2 (27483.8 y_t^4 - 52858 y_t^2 g_3^2 + 18215.3 g_3^4) \nl
& + \lambda y_t^2 (-51279 y_t^6 - 5139.56  y_t^4 g_3^2 + 50795.3 y_t^2 g_3^4 - 33858.8 g_3^6) \nl
& y_t^4 (-24318.2 y_t^6+ 72896  y_t^4 g_3^2 - 73567.3 y_t^2 g_3^4 + 36376.5 g_3^6) \Bigg\} L^2 . \label{eqn:d4lambda}
}
To simplify the expression, we have excluded the $y_b, y_\tau$ contributions beyond one-loop order, and $g_1, g_2$ contributions beyond two-loop order, although they propagate at higher orders in terms that include $\b_\l^{(1)}, \b_{y_t}^{(1)}$. 

 We use two different calculations of the values of the SM parameters at the renormalization scale $Q=M_S$. In the simpler, approximate calculation, using Eq.~(\ref{eqn:lambdaQlow}), $g_3(M_S)$ and $y_t(M_S)$ are computed from $g_3(M_t)$ and $y_t(M_t)$ using two- and three-loop fixed-order formulae, respectively: %since they appear at one- and two-loop in $\lambda$:
\ba{
y_t (M_S) &= y_t + \k \Bigg\{ \b_{y_t}^{(1)} L 
+ \frac{\b_{y_t}^{(1,1)}}2 L^2 + \frac{\b_{y_t}^{(1,2)}}{3!} L^3 \Bigg\} 
+ \k^2 \Bigg\{ \b_{y_t}^{(2)} L + \frac{\b_{y_t}^{(2,1)}}2 L^2 \Bigg\} 
+ \k^3 \b_{y_t}^{(3)} L , \label{eqn:ytMS} \\
g_3 (M_S) &= g_3 + \k \Bigg\{ \b_{g_3}^{(1)} L + \frac{\b_{g_3}^{(1,1)}}2 L^2 \Bigg\}
+ \k^2 \b_{g_3}^{(2)} L . \label{eqn:g3MS}
}
Parameters on the right-hand sides of Eqs. (\ref{eqn:ytMS}) and (\ref{eqn:g3MS}) are evaluated at $M_t$, and the $\b$ functions are given in Appendix \ref{app:betafuncs}.  $\l (M_S)$ is computed using Eq. (\ref{eqn:lambdaMSSM}), with $y_t(M_S)$ and $g_3(M_S)$ appearing in Eqs. (\ref{eqn:hytMSbar}-\ref{eqn:Dhb}, \ref{eqn:2loopasat}) obtained from Eqs. (\ref{eqn:ytMS}) and (\ref{eqn:g3MS}). In Eqs.~(\ref{eqn:hybMSbar}-\ref{eqn:hytauMSbar}) only, we perform a one-loop fixed-order running with couplings at $M_t$ to approximate $y_b$ and $y_\tau$ at $M_S$:
\ba{
y_b (M_S) &= y_b (M_t) \Big[1 + \k \Big( \frac32 y_t^2 - 8 g_3^2 - \frac94 g_2^2 - \frac14 g_1^2 \Big) L \Big] , \\
y_\tau (M_S) &= y_\tau (M_t) \Big[1 + \k \Big( 3 y_t^2 - \frac94 g_2^2 - \frac94 g_1^2 \Big) L \Big] .
}
In the tree-level $\lambda_{\text{tree}}$ (Eq.~(\ref{eqn:lambdatree})) of the zeroth-order $\lambda(M_S)$, i.e. the first term on the right-hand side of Eq.~(\ref{eqn:lambdaanalytic}), we have also approximated $g_1$ and $g_2$ at $M_S$ using a one-loop fixed-order running:
\ba{
g_1^2 (M_S) &= g_1^2 (M_t) \Big[ 1 + 2\k \Big( \frac{41}{10} g_1^2 L + \frac25 g_1^2 L_\m \Big) \Big] , \\
g_2^2 (M_S) &= g_2^2 (M_t) \Big[1 + 2\k \Big( -\frac{19}6 g_2^2 L + 2 g_2^2 L_\m \Big) \Big] .
}
Elsewhere in the calculation for $\l(M_S)$ and in Eqs.~(\ref{eqn:d1lambda}-\ref{eqn:d4lambda}), we use the $Q = M_t$ values for $g_1, g_2, y_b, y_\tau$. 

To convert the running mass into the pole mass, we use the one-loop formula 
\ba{
\label{eqn:selfenergies}
M_h^2 &= \l (M_t) v^2 (M_t) + \k \Bigg\{ 
3y_t^2 (4\mb_t^2 - m_h^2) B_0(\mb_t, \mb_t, m_h) - \frac92 \l m_h^2 \Big[2 - \frac{\pi}{\sqrt3} - \log \frac{m_h^2}{Q^2} \Big] \nl
& \qquad - \frac{v^2}4 \Big[ 3g_2^4 - 4\l g_2^2 + 4\l^2 \Big] B_0(m_W, m_W, m_h) \\
& \qquad - \frac{v^2}8 \Big[ 3(g_2^2 + g_Y^2)^2 - 4\l (g_2^2 + g_Y^2) + 4\l^2 \Big] B_0(m_Z, m_Z, m_h) \nl
& \qquad + \frac12 g_2^4 \Big[ g_2^2 - \l \Big( \log\frac{m_W^2}{Q^2} - 1 \Big) \Big] + \frac14 (g_2^2 + g_Y^2) \Big[ (g_2^2 + g_Y^2)) - \l \Big( \log\frac{m_Z^2}{Q^2} - 1 \Big) \Big]
\Bigg\} , \notag
}
where $B_0$ is the one-loop Passarino-Veltman integral
\be{
B_0(m_1, m_2, m_3) = -\int_0^1 \log \Big[ \frac{(1-x)m_1^2 + x m_2^2 - x(1-x) m_3^2}{Q^2} \Big],
}
and all quantities appearing at one-loop are $\MSbar$ running parameters with $Q = M_t$. This correction is a small effect, of order 0.5 GeV.

Together with the threshold corrections given in Sec.~\ref{sect:matching}, Eqs.~(\ref{eqn:mhrun})-(\ref{eqn:selfenergies}) can be used to compute the Higgs mass to four-loop NNLL accuracy, in the approximation that one scale controls the MSSM scalar and gluino masses and a second (possibly equal) scale controls the electroweakino masses. We will compare these analytic formulas with the results from numerically integrating the RGEs, to understand the regimes in which the fixed-order calculation is good.

% FIGURE: XtMS 0, \tan\beta = 20
\begin{figure}[tb]
\begin{center}
\includegraphics[scale=0.625]{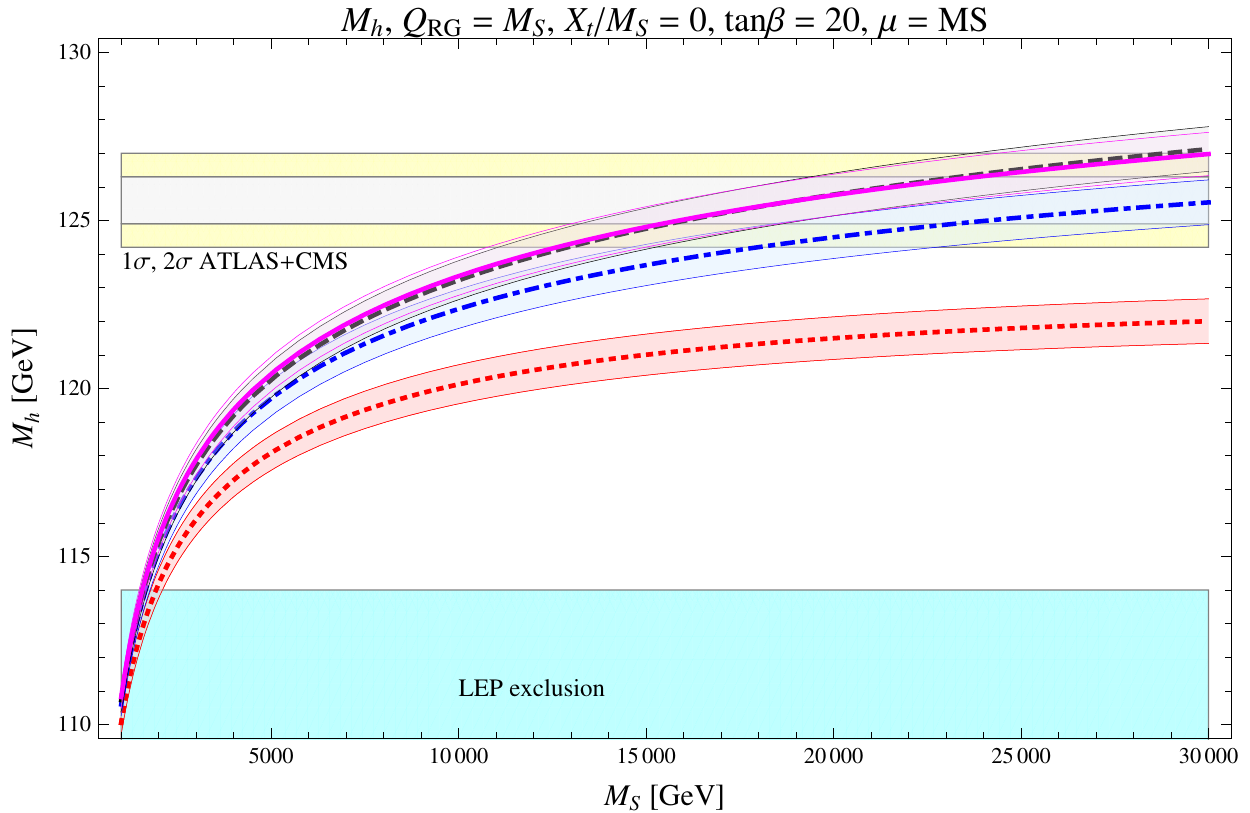}
\includegraphics[scale=0.625]{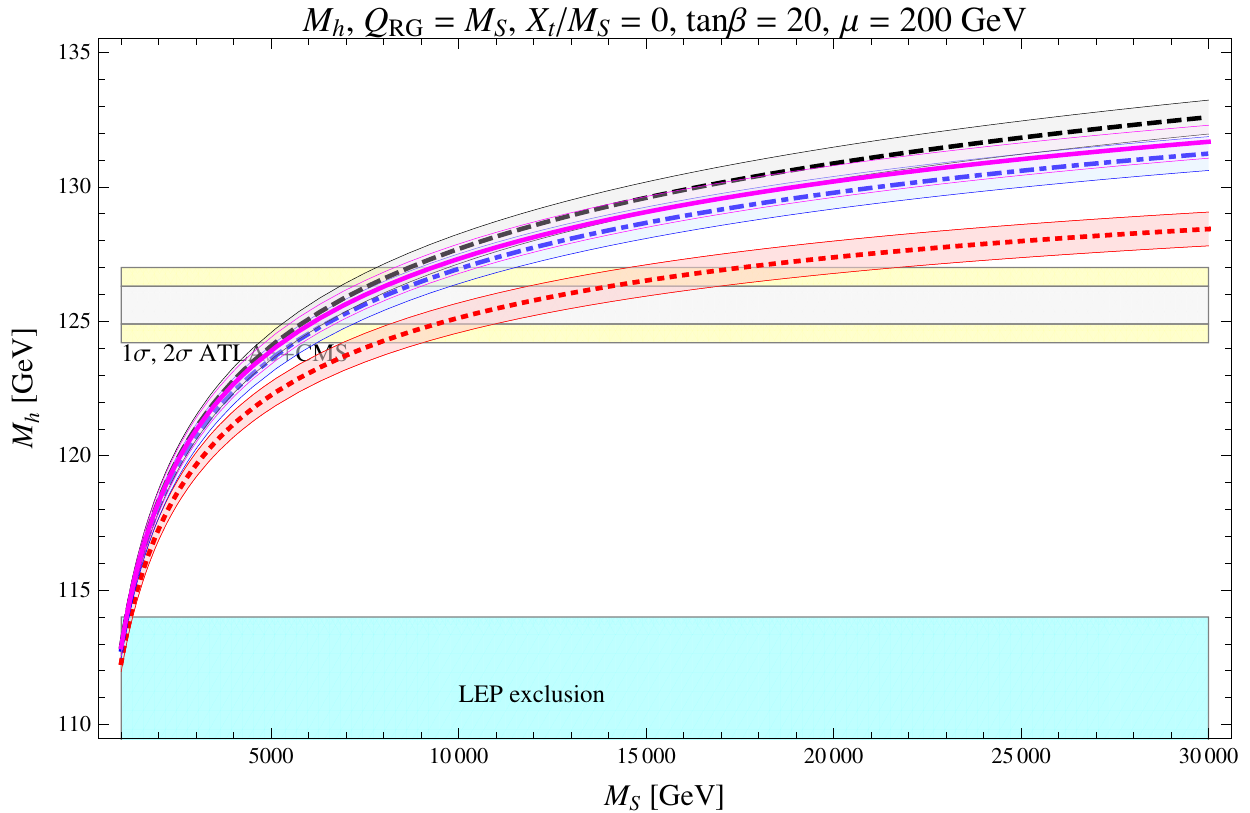}
\includegraphics[scale=0.625]{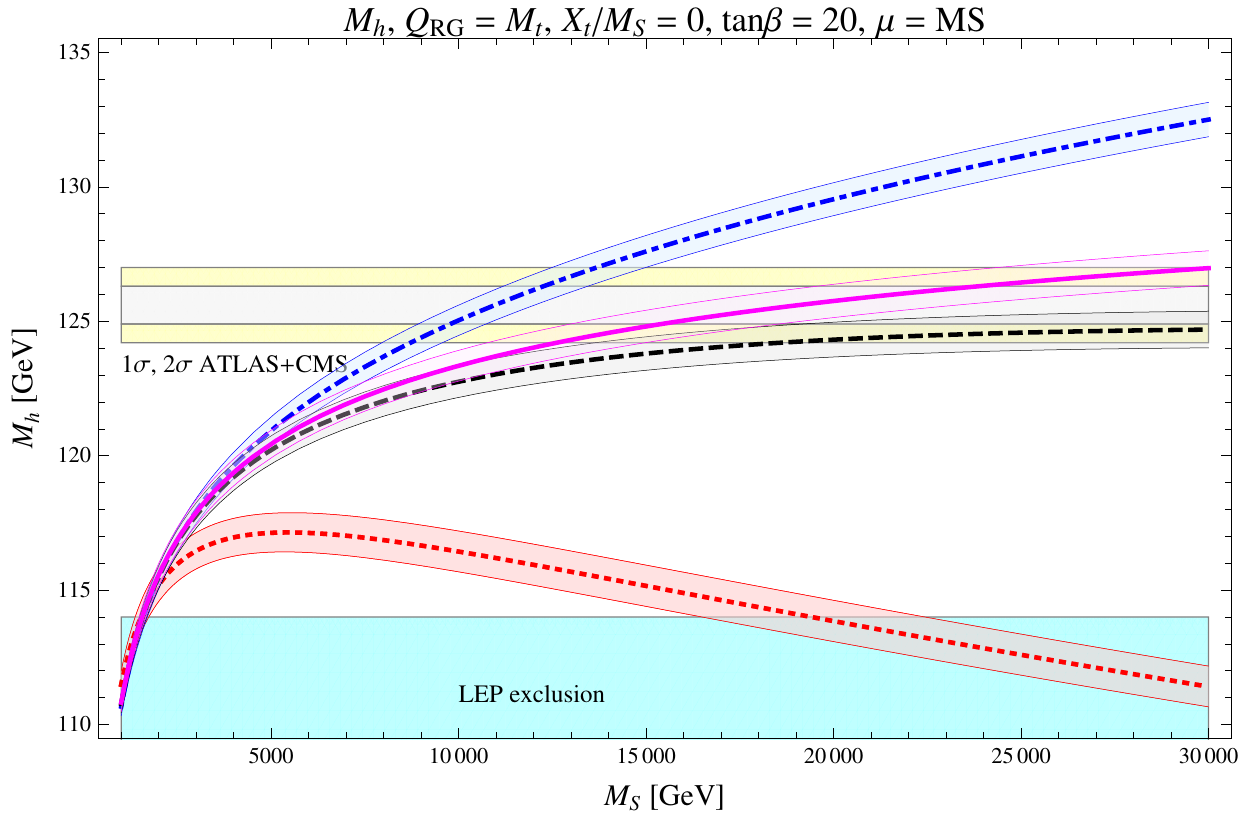}
\includegraphics[scale=0.625]{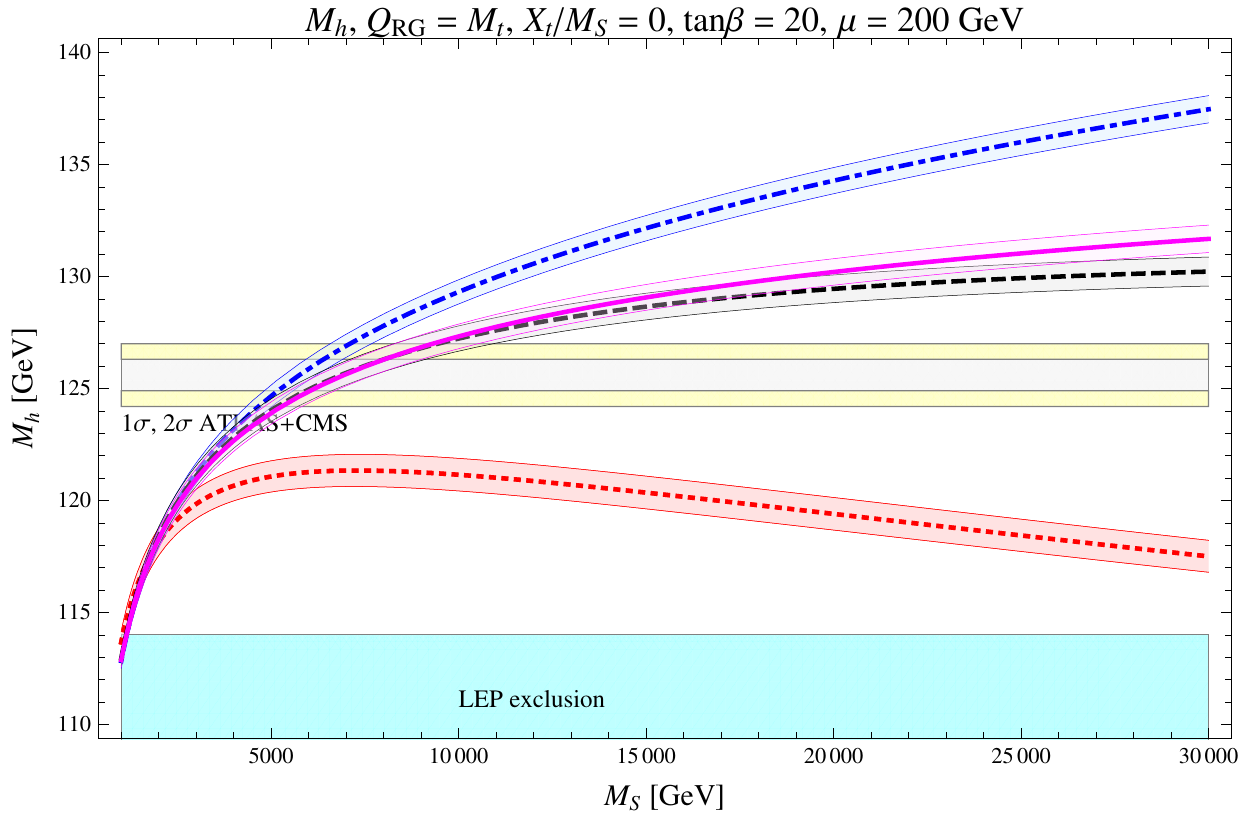}
\end{center}
\caption{Plots of Higgs mass $M_h$ versus the SUSY scale $M_S$ for $\XtMS = 0, \tan\b = 20$ with $\mu = M_S$ (left column) and $\mu = 200$ GeV (right column). The solid magenta, black dotted, blue dot-dashed, and red dotted lines correspond to the resummed calculation and the four-, three-, and two-loop fixed-order calculations, respectively. The shaded regions for each calculation indicate the uncertainty from varying $M_t$ by the $1\sigma$ values. The top (bottom) figure in each column corresponds to the fixed-order calculation for $Q = M_S$ $(Q = M_t)$. The grey (yellow) region corresponds to the approximate $1\sigma$ $(2\sigma)$ values for the Higgs mass $M_h \sim 125.6 \pm 0.7$ GeV measured by the ATLAS and CMS collaborations, and the cyan region is excluded by LEP.}
\label{fig:XtMS0tanb20}
\end{figure}

% FIGURE: XtMS max, \tan\beta = 4
\begin{figure}[tb]
\begin{center}
\includegraphics[scale=0.625]{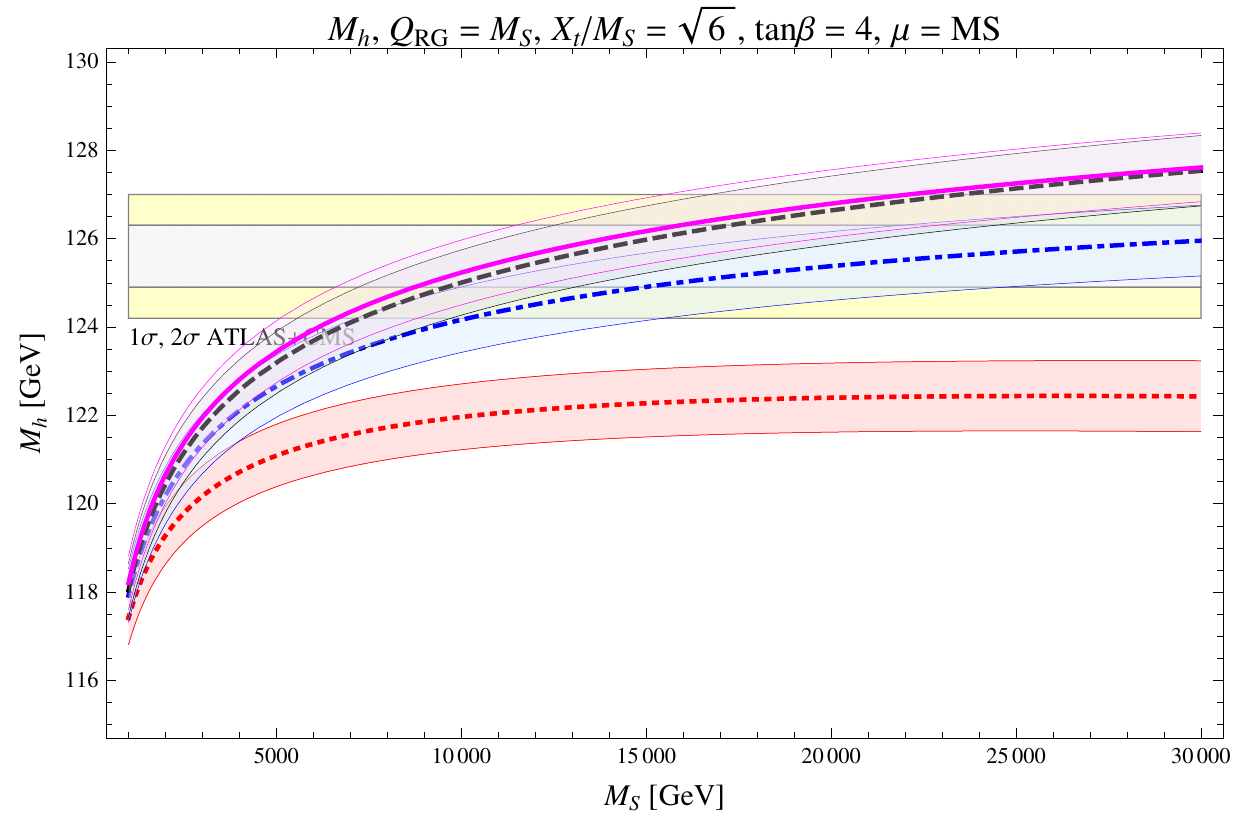}
\includegraphics[scale=0.625]{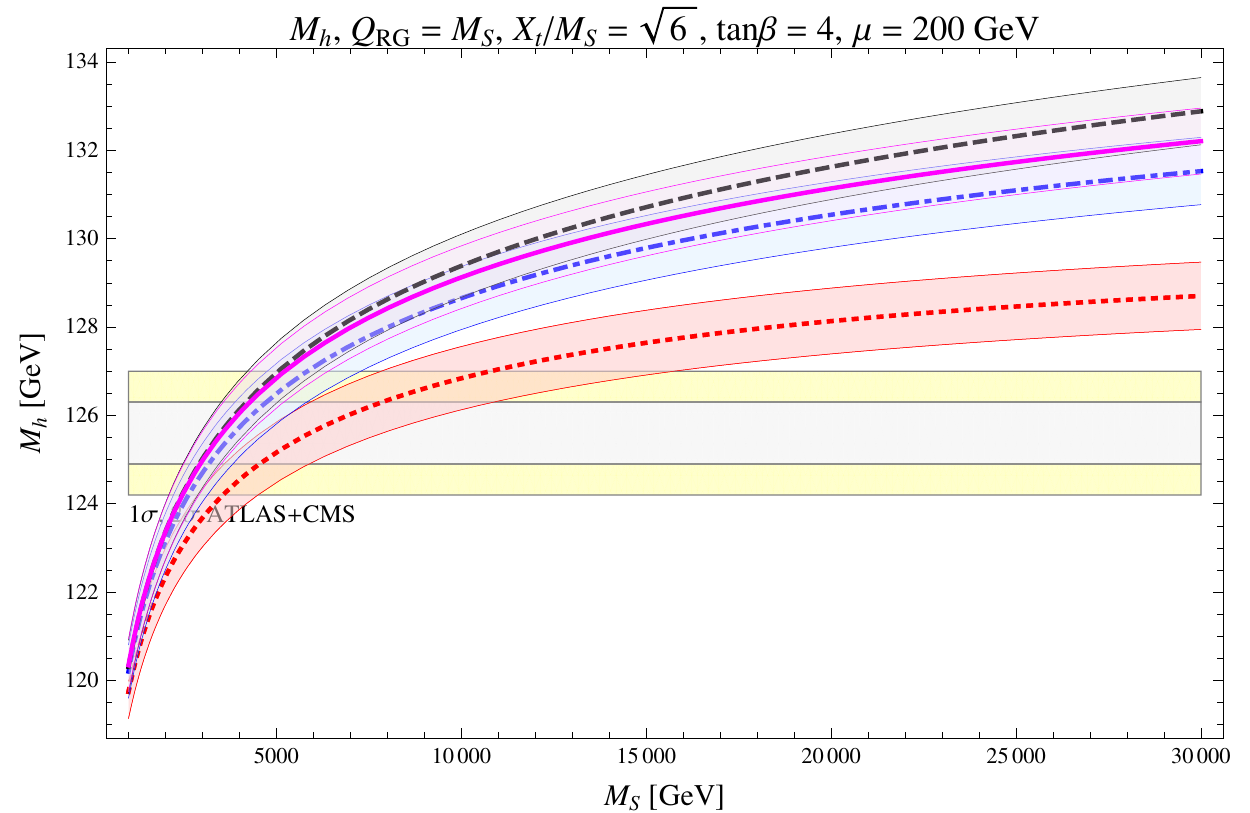}
\includegraphics[scale=0.625]{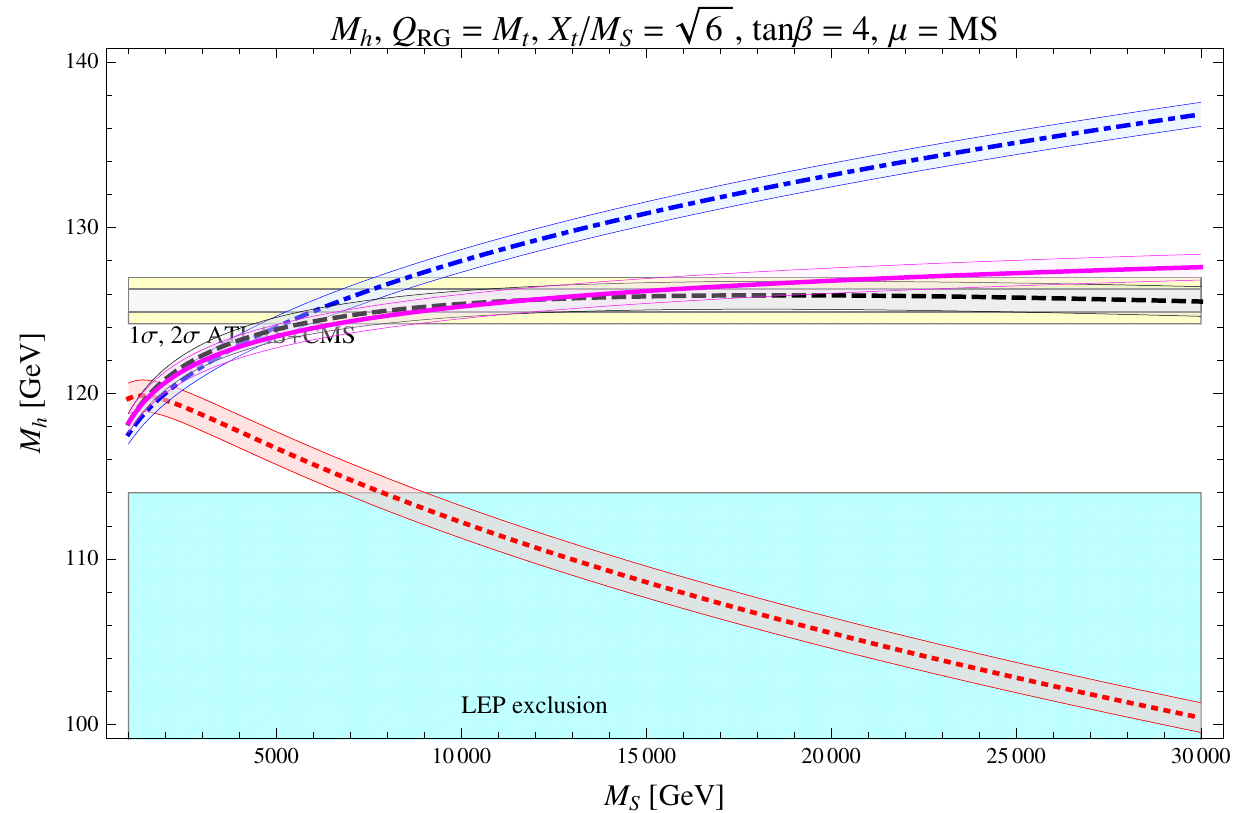}
\includegraphics[scale=0.625]{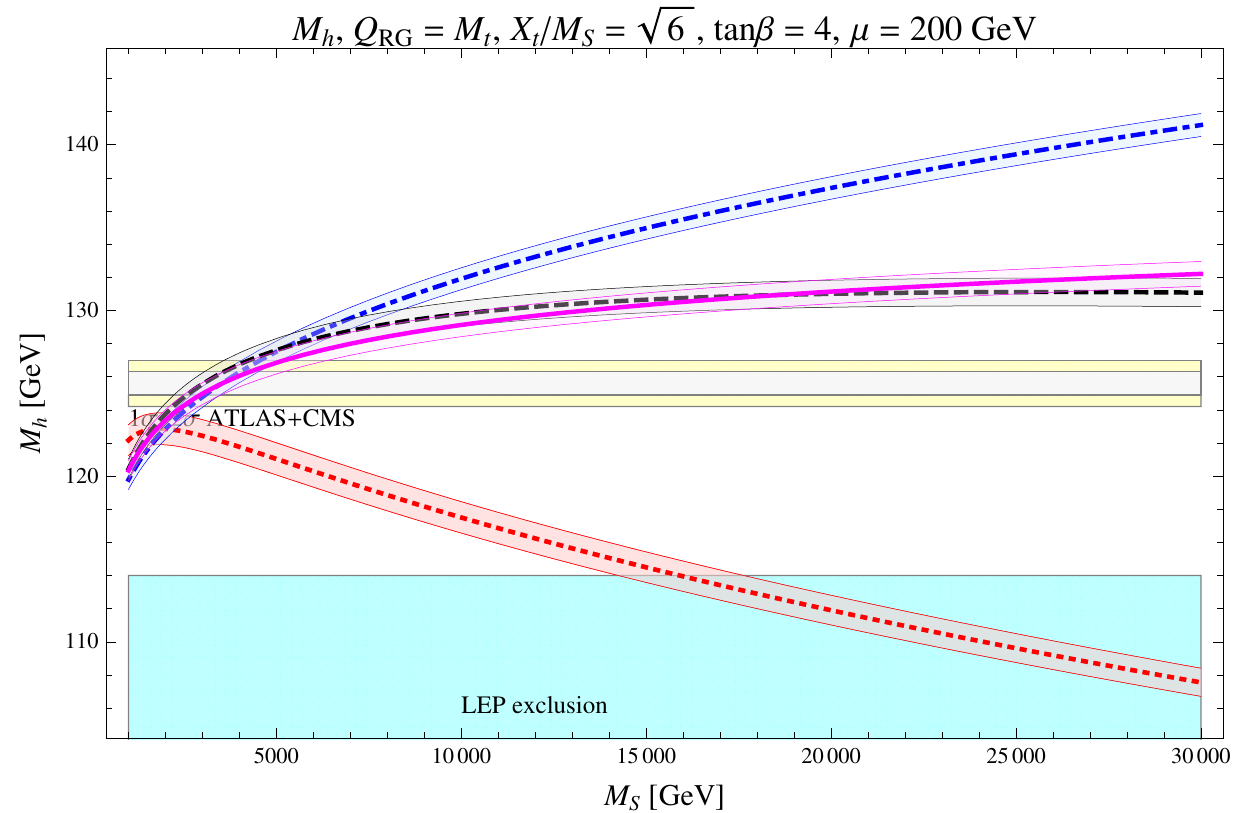}
\end{center}
\caption{Plots of Higgs mass $M_h$ versus the SUSY scale $M_S$ for $\XtMS = \sqrt6, \tan\b = 4$ with $\mu = M_S$ (left column) and $\mu = 200$ GeV (right column). See Fig. \ref{fig:XtMS0tanb20} for details.}
\label{fig:XtMSmaxtanb4}
\end{figure}

To begin the comparison, we plot $M_h$ in Figs. \ref{fig:XtMS0tanb20}~and~\ref{fig:XtMSmaxtanb4} corresponding to two scenarios with the values 
\begin{enumerate}
\item $\tan\b = 20, \XtMS = 0$, and
\item $\tan\b = 4, \XtMS = \sqrt{6}$, 
\end{enumerate}
and we consider the range of $M_S$ between 1 and 30 TeV. These figures include results for the resummed calculation and the fixed-order calculations at two-loop, three-loop, and four-loop with couplings evaluated at $Q = M_S$ and $Q = M_t$. 

For the two-loop fixed-order calculation with couplings at $Q = M_t$, we have used the NLO value for $y_{t, \text{NLO}} (M_t) = 0.95096$, whereas all other calculations use the NNLO value $y_{t, \text{NNLO}} (M_t) = 0.93697$ from Eq.~(\ref{eqn:ytNNLO}). This is responsible for the disagreement between the two-loop curve and the other curves in the $Q = M_t$ plots in Figs. \ref{fig:XtMS0tanb20}~and~\ref{fig:XtMSmaxtanb4} for low $M_S \sim 1$~TeV, where the threshold corrections are more important. 

We observe that the $Q = M_S$ fixed-order results converge approximately monotonically with increasing loop-order towards the resummed result, whereas the $Q = M_t$ exhibits the alternating behaviour and shows significantly worse agreement for large $M_S \geq 10$ TeV. The resummed method and the $Q = M_S$ four-loop fixed-order calculation differ by less than 0.5 GeV in the $\mu = M_S$ case, and by just over 1 GeV in the $\mu = 200$ GeV case; the difference between the resummed and three-loop results is less than 1.5 GeV and 1 GeV, respectively. The value of the pole mass $M_t$ is the dominant source of parametric uncertainty for $M_h$: taking the $1\sigma$ high and low values for $M_t$ changes $M_h$ by about 0.8 GeV. We find that to achieve $M_h \sim 125.6$ GeV with $\mu = 200$ GeV, a SUSY scale of $M_S \sim 7$ (3.5) TeV is required in scenario 1 (2); for $\mu = M_S$, we require $M_S \sim 18$ (12) TeV for scenario 1 (2). For $\tan\beta = 30$ and $\XtMS = \sqrt{6}$, we find $M_S \sim 1.5$ (1) TeV for $\mu = M_S$ (200 GeV).

% FIGURE: Q = M_S RK, parameters as in previous two figures
\begin{figure}[tb]
\begin{center}
\includegraphics[scale=0.625]{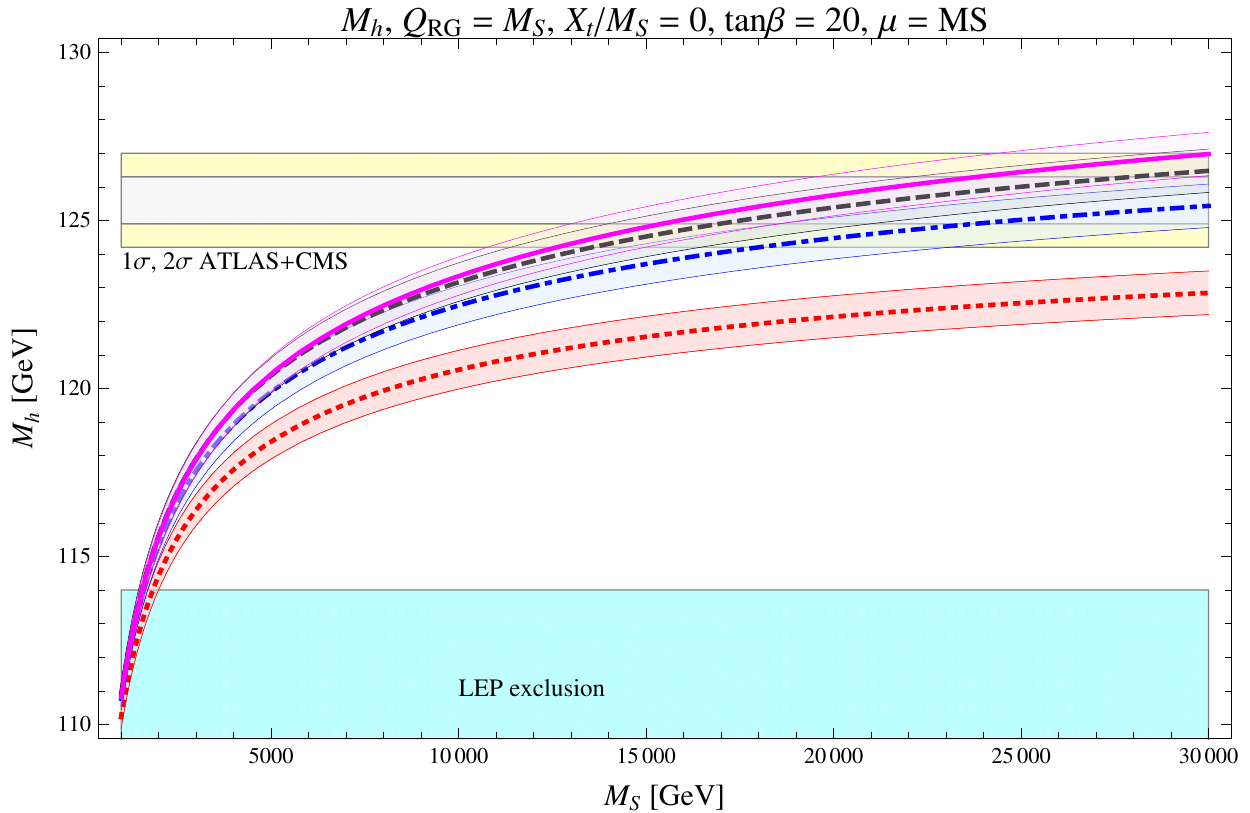}
\includegraphics[scale=0.625]{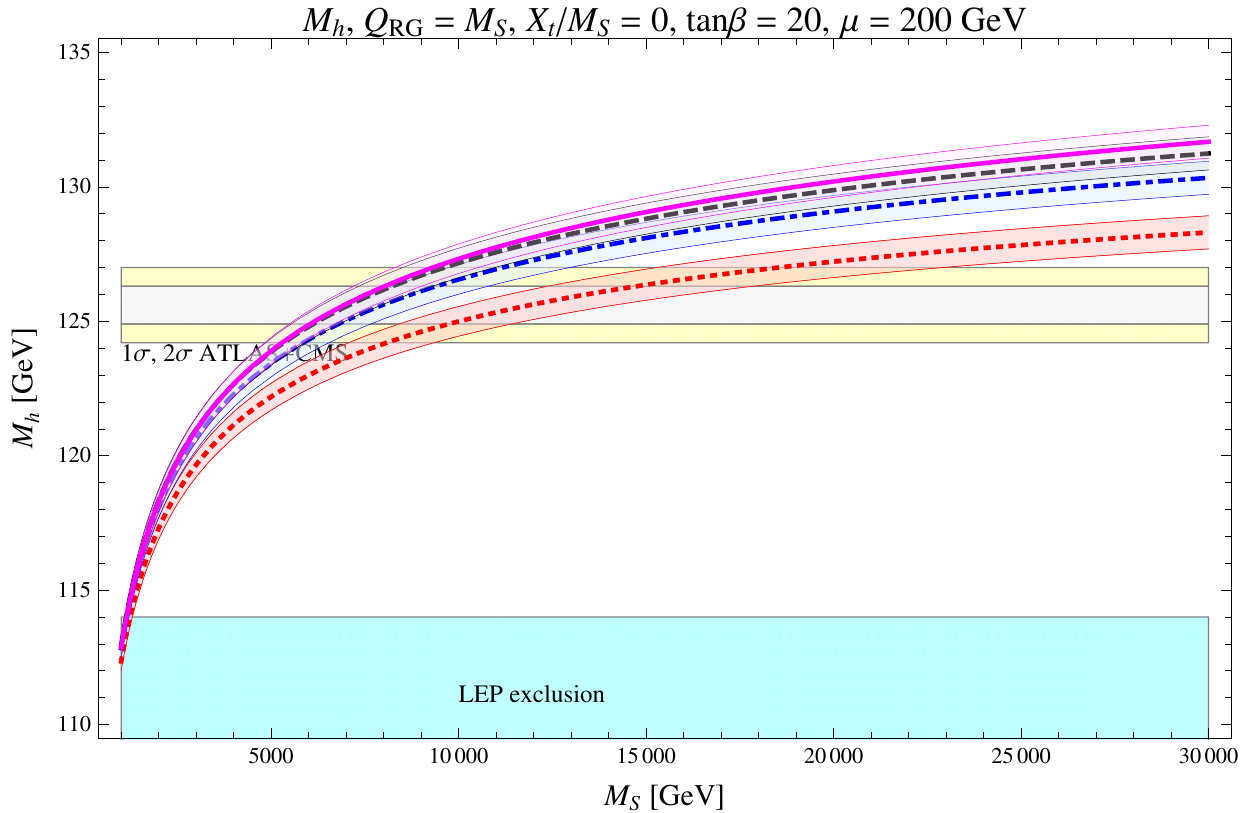}
\includegraphics[scale=0.625]{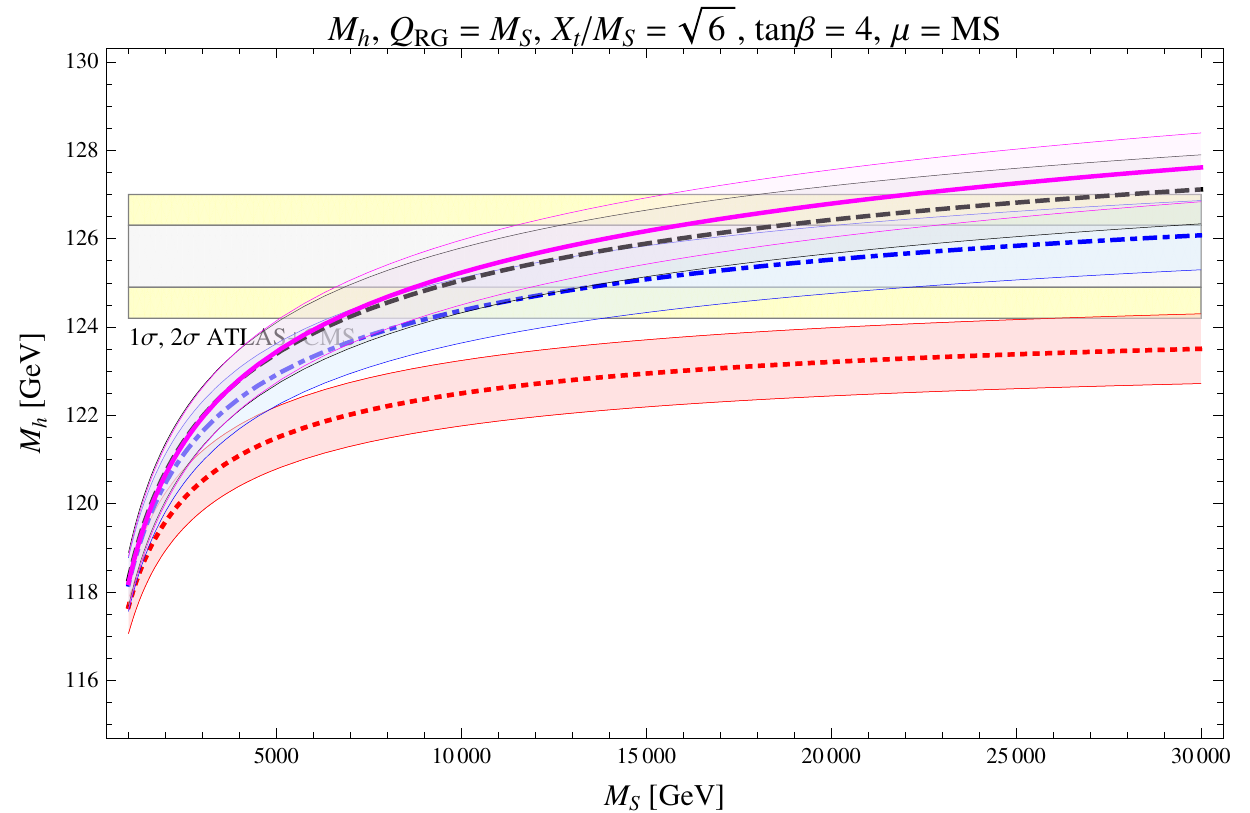}
\includegraphics[scale=0.625]{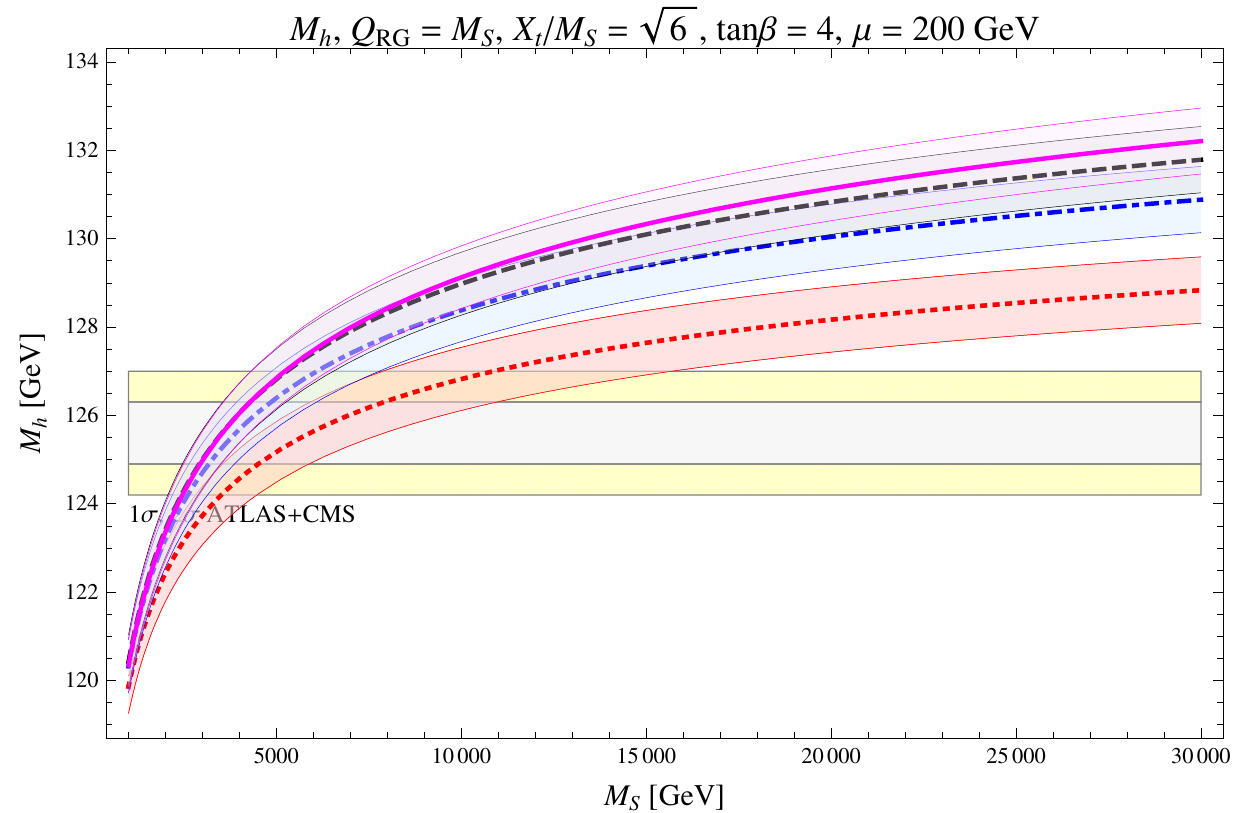}
\end{center}
\caption{Plots of Higgs mass $M_h$ versus the SUSY scale $M_S$ using the fixed-order calculation with couplings at $M_S$ obtained from the full numerical integration. We use the values $\XtMS = 0, \tan\b = 20$ (top row) and $\XtMS = \sqrt6, \tan\b = 4$ (bottom row), with $\mu = M_S$ (left column) and $\mu = 200$ GeV (right column). See Fig. \ref{fig:XtMS0tanb20} for details.}
\label{fig:hybridRK}
\end{figure}

As mentioned above, we have also performed a second fixed-order calculation, differing in the values taken for the running parameters at $Q=M_S$. In the second case, we use the exact running parameters, amounting to a hybrid calculation, since they are extracted from the same numerical integration algorithm used to perform the fully resummed computation of $m_h$. The results for the two scenarios above are shown in Fig. \ref{fig:hybridRK}. As should be expected, the analytic approximation now converges monotonically to the resummed result, and the four-loop result remains within 0.5 GeV of the resummed result for both $\mu = M_S$ and $\mu = 200$ GeV in both scenarios. The difference between the resummed and three-loop results is roughly 2 to 3 times greater, between 1.2 and 1.5 GeV.

From these plots we conclude that the four-loop NNLL result with $Q=M_S$ is equal to the resummed result, within the current top mass uncertainties, for $M_S$ as large as tens of TeV. Unsurprisingly, the three-loop result diverges more rapidly, and underestimates the Higgs mass in the case $Q=M_S$. 

\begin{figure}[tb]
\begin{center}
\includegraphics[scale=0.625]{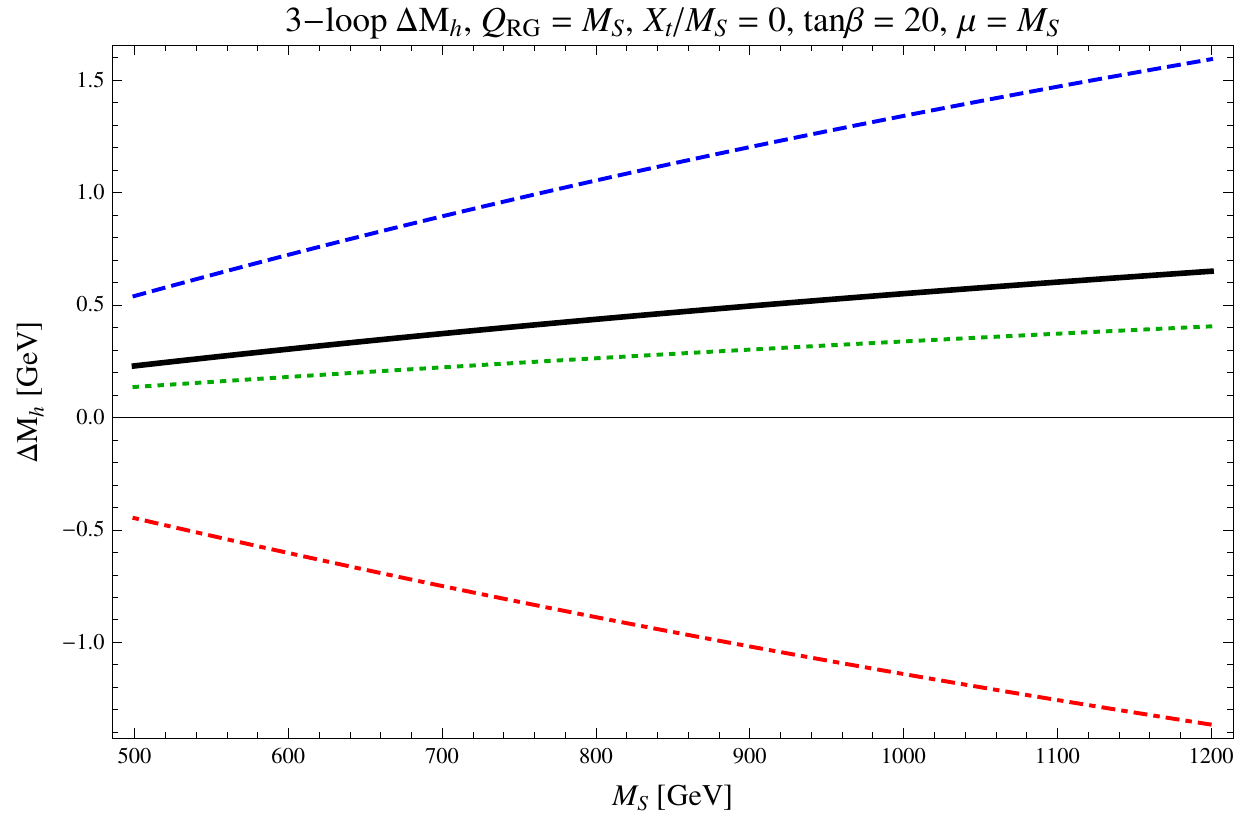}
\includegraphics[scale=0.625]{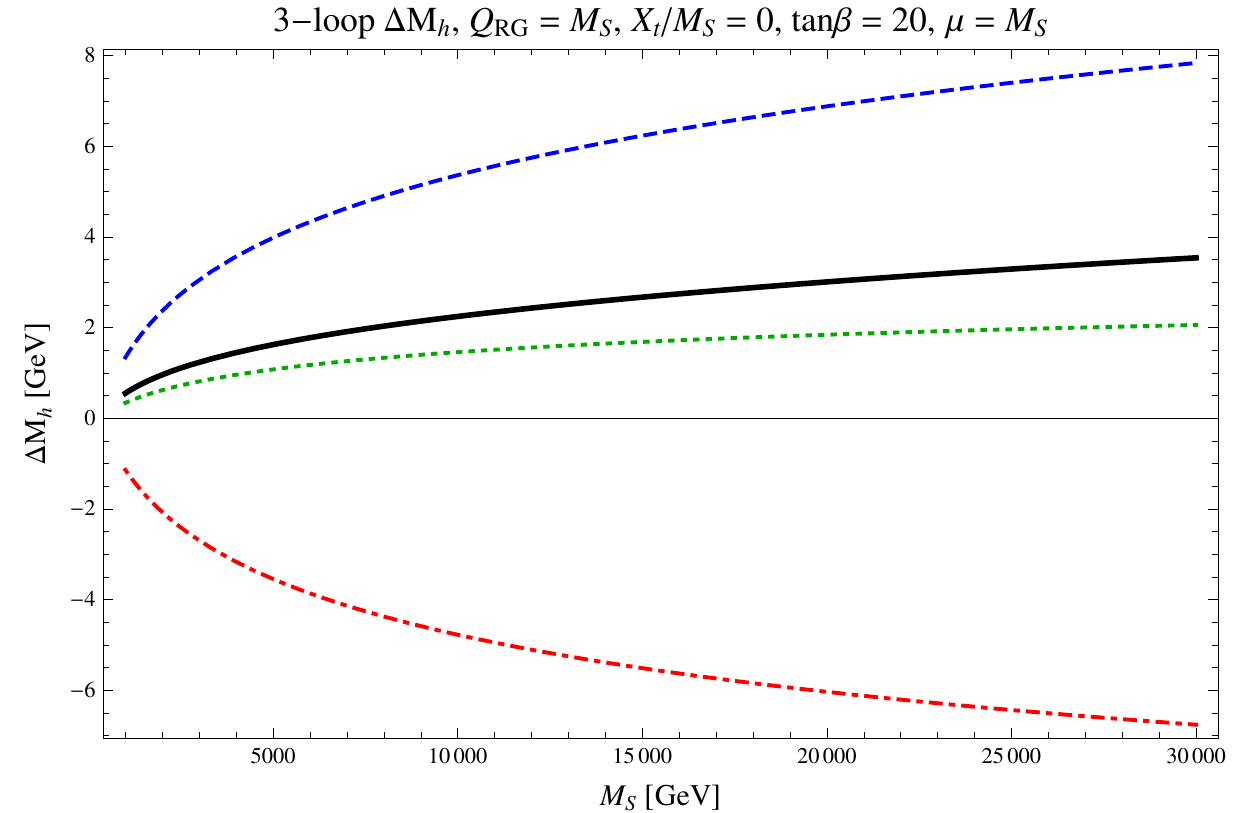}
\includegraphics[scale=0.625]{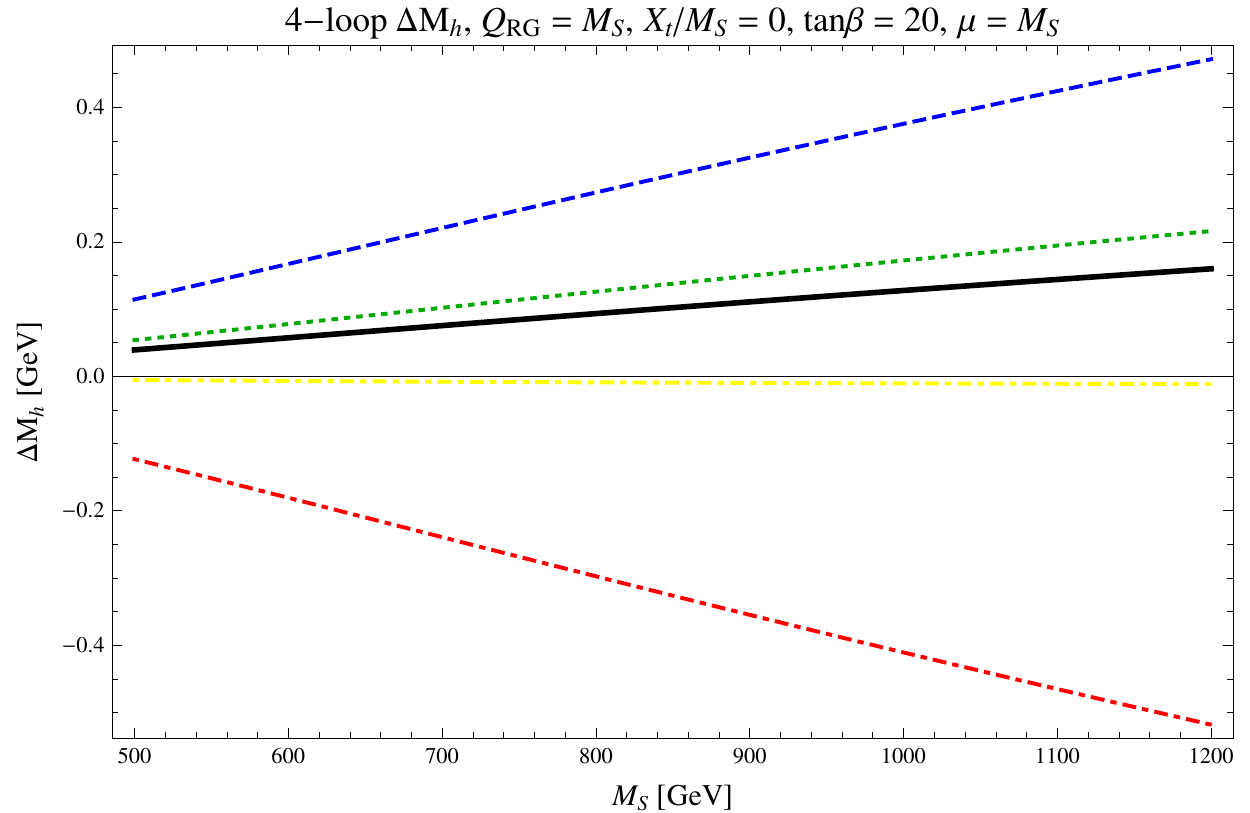}
\includegraphics[scale=0.625]{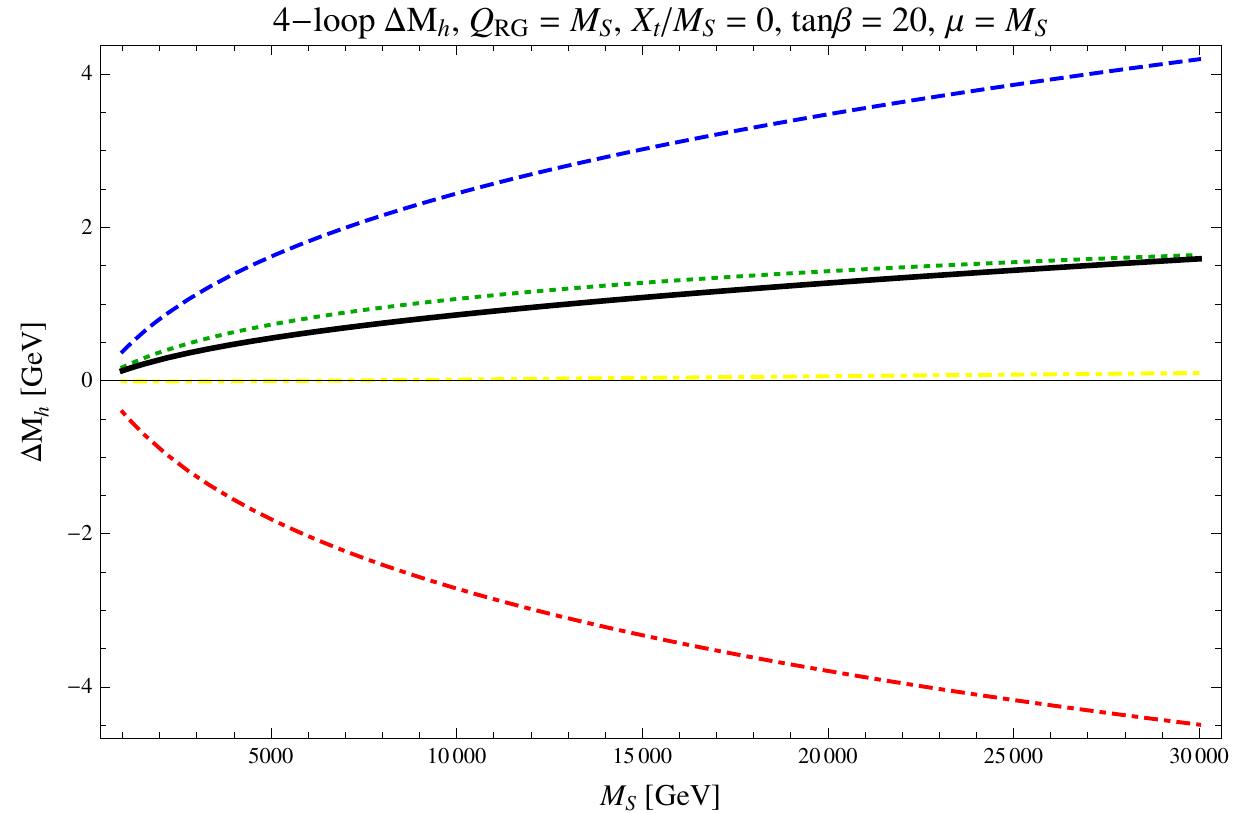}
\end{center}
\caption{Plots of the separate contributions of terms at $n$-loop order proportional to $g_3^{2k}, 0 \leq k \leq n-1$. For the three-loop figures (top row), the blue dashed (red dot-dashed) lines include the terms proportional to $g_3^4 y_t^4$ $(g_3^2 y_t^6)$. The green dotted line is the remainder, and the black solid line is the total difference from the two-loop result. Similarly, for the four-loop figures (bottom row), the blue dashed (red dot-dashed, green dotted) lines include the terms proportional to $g_3^6 y_t^4$ $(g_3^4 y_t^6, g_3^2 y_t^8)$, and the yellow dotted line is the remainder. }
\label{fig:g3cancellation}
\end{figure}

On the other hand, it is also possible to overestimate corrections to the Higgs mass by considering only a subset of the three-loop terms. This is due to a striking accidental cancellation at leading log in $\d_3 \l$ ($\d_4 \l$) between leading $g_3^4 y_t^4$ $(g_3^6 y_t^4)$ and subleading $g_3^2 y_t^6$ and $y_t^8$ $(g_3^4 y_t^6, g_3^2 y_t^8$, and $y_t^{10})$ contributions; these are the last three (four) terms before the large closing curly braces in Eqs.~(\ref{eqn:d3lambda})~and~(\ref{eqn:d4lambda}). We note that the cancellation persists to a lesser degree at each subleading log order in $L^k$. The cancellation at leading log was first noted in \cite{Martin:2007pg}, the result of which we extend to higher values of $M_S$ and improve by including subleading log corrections. Our result is exhibited in Fig. \ref{fig:g3cancellation}. Although the individual contributions to the radiative corrections are about 50\% larger in magnitude than was found in \cite{Martin:2007pg}, our cancellation is more efficient, in part because we are using higher values for $M_t$ and $M_h$ and have included subleading log orders.

Figure \ref{fig:g3cancellation} raises the concern that a partial three-loop fixed-order computation that includes only $g_3^4 y_t^4$ corrections and not $g_3^4 y_t^6$ terms may overestimate the Higgs mass by several GeV for $M_S$ of order 10 TeV. This may explain in part the discrepancy between the required stop scales found with resummation and those found in the analysis of~\cite{Feng:2013tvd}.

% FIGURE: g_3^{2k} CANCELLATIONS

% FIGURE: tanb v M_S
\begin{figure}[tb]
\begin{center}
\includegraphics[scale=0.58]{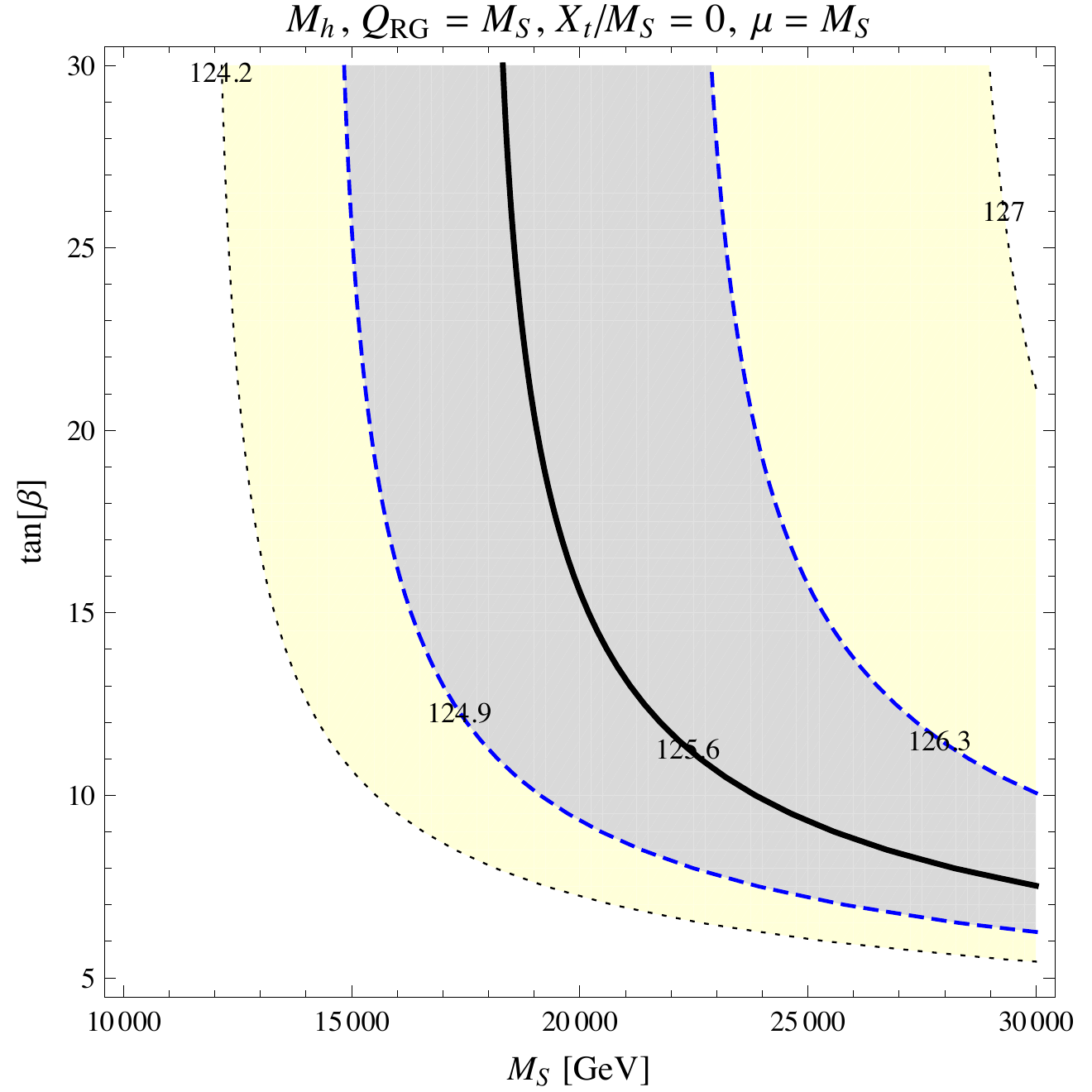} \quad
\includegraphics[scale=0.58]{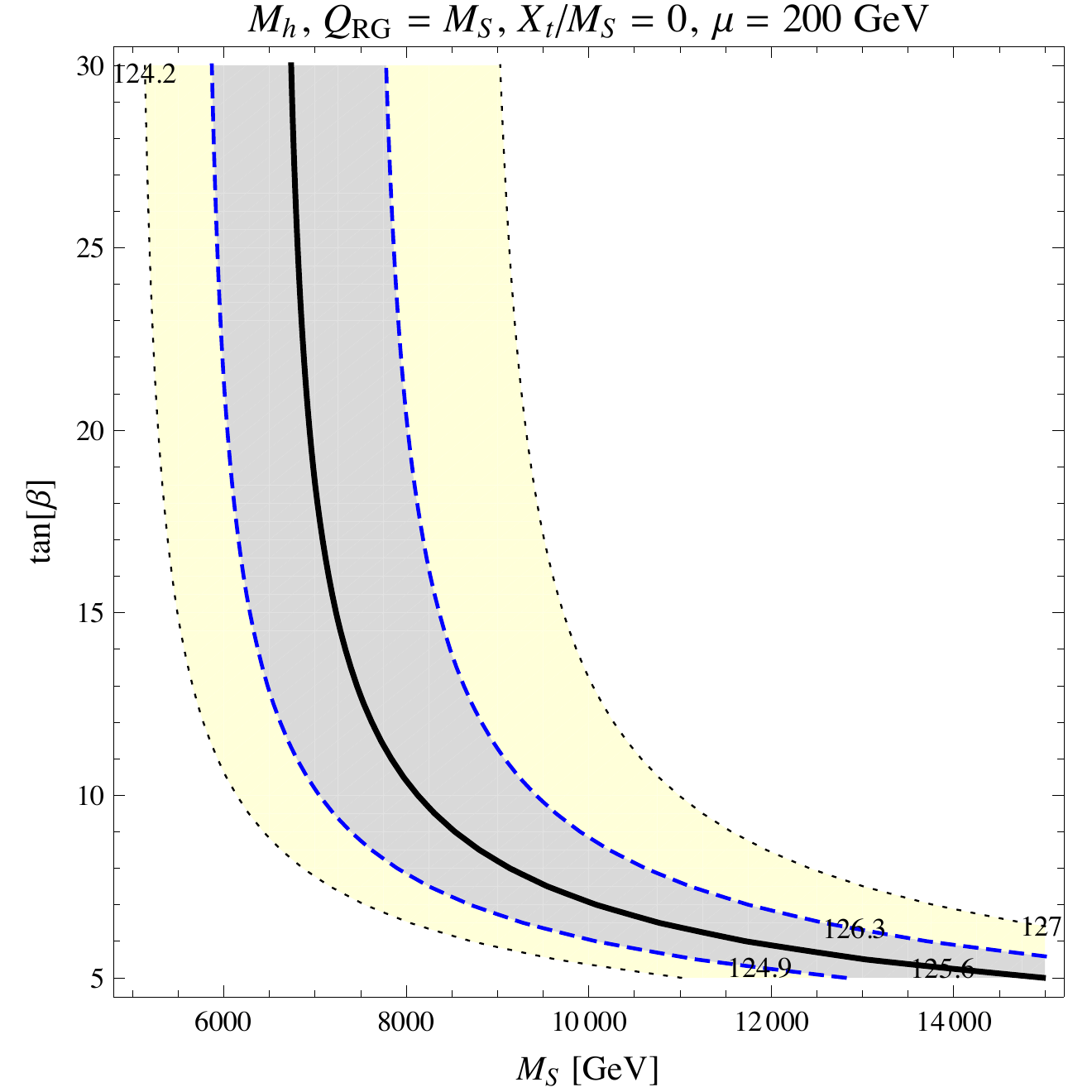}
\includegraphics[scale=0.58]{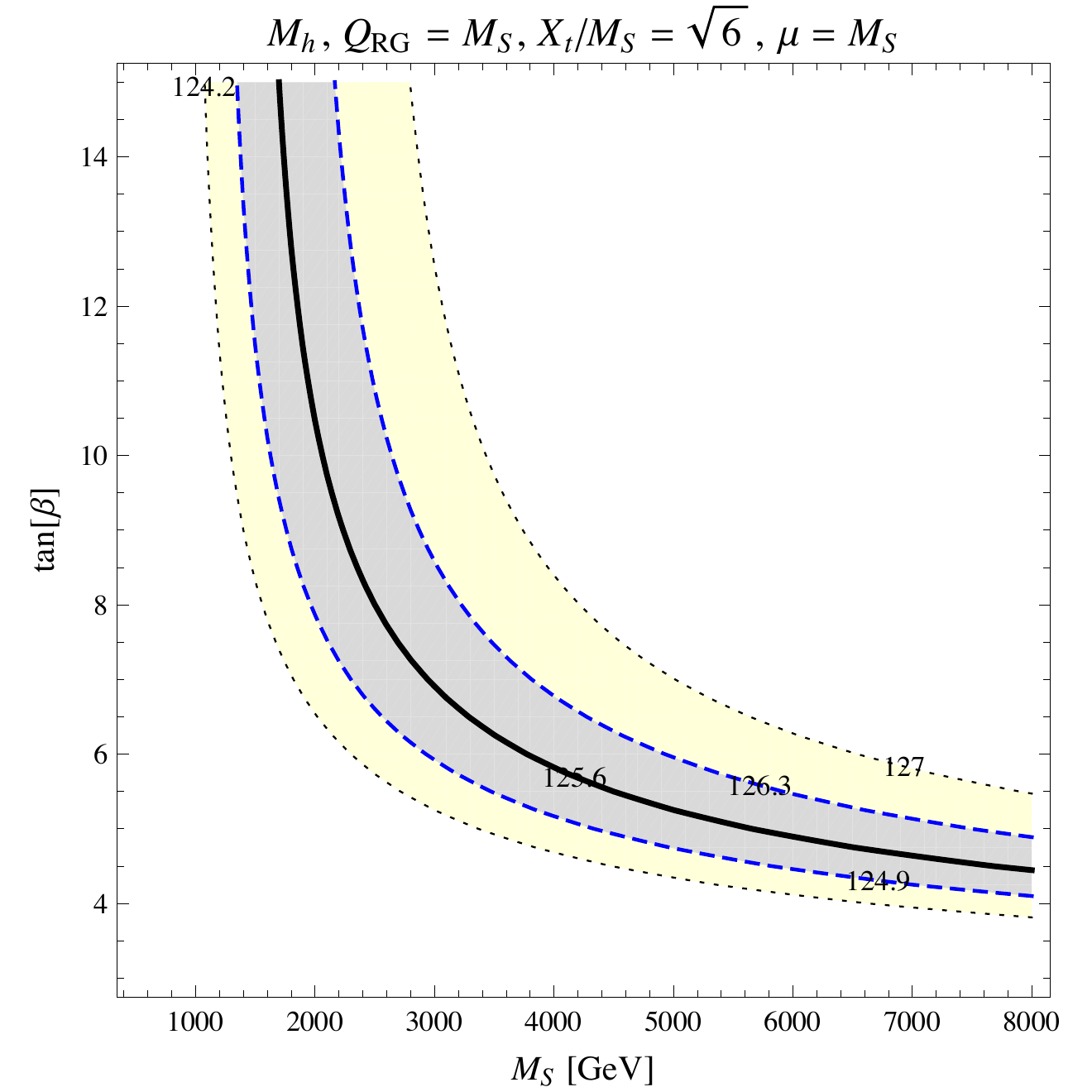} \quad
\includegraphics[scale=0.58]{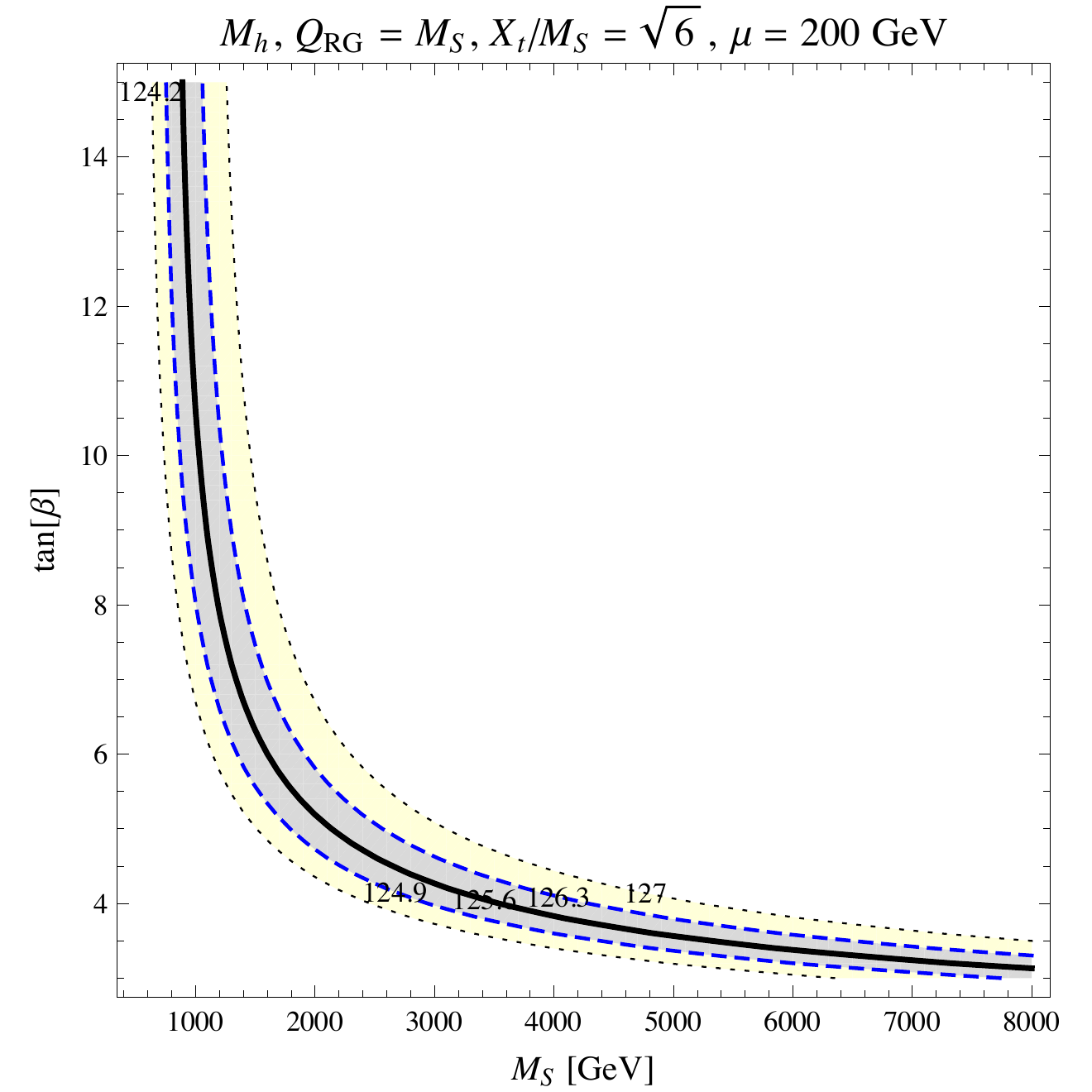}
\end{center}
\caption{Plots of central (solid), $1\sigma$ (dashed), and $2\sigma$ (dotted) contours of the Higgs mass $M_h$ in the $\tan\b$ vs. $M_S$ plane for values of $\XtMS = 0, \sqrt6$ (top, bottom rows) and $\mu = M_S$, 200 GeV (left, right columns).}
\label{fig:tanbvMS}
\end{figure}

In Fig. \ref{fig:tanbvMS}, we show contours of the central, $1\sigma$, and $2\sigma$ values for $M_h$ in the $(M_S, \tan\beta)$ plane for $\XtMS = 0, \sqrt6$ and $\mu = M_S$, 200 GeV. For $\XtMS = 0$ and $\mu = M_S$ (200 GeV), we see again that for large $\tan\beta > 20$, we require $M_S \sim 18$ (7) TeV to achieve $M_h \sim 125.6$ GeV, although within uncertainties, this scale can vary by a few TeV. For a fixed value of moderate to large $\tan\beta \gtrsim 10$, the relatively large spread in $M_S$ required to obtain $M_h\sim 125.6 \pm 0.7$ GeV corresponds to the shallow slope of $M_h$ in Fig. \ref{fig:XtMS0tanb20} at large $M_S$; the central value, however, constrains $M_S$ to the range $18 \TeV \lesssim M_S \lesssim 24$ TeV $(6.5 \TeV \lesssim M_S \lesssim 8$ TeV).

For maximal mixing, $M_h$ greatly constraints the parameter space. The central value favours $M_S <  2$ (1) TeV for $\tan\beta > 10$ for $\mu = M_S$ (200 GeV). Here, we again see the larger spread in $M_S$ at low $\tan\beta$. As in the case for zero mixing, this allowed range of a few TeV can be mapped to the equivalent shallow slope in Fig. \ref{fig:XtMSmaxtanb4}.

% FIGURE: M_h v XtMS
\begin{figure}[tb]
\begin{center}
\includegraphics[scale=0.75]{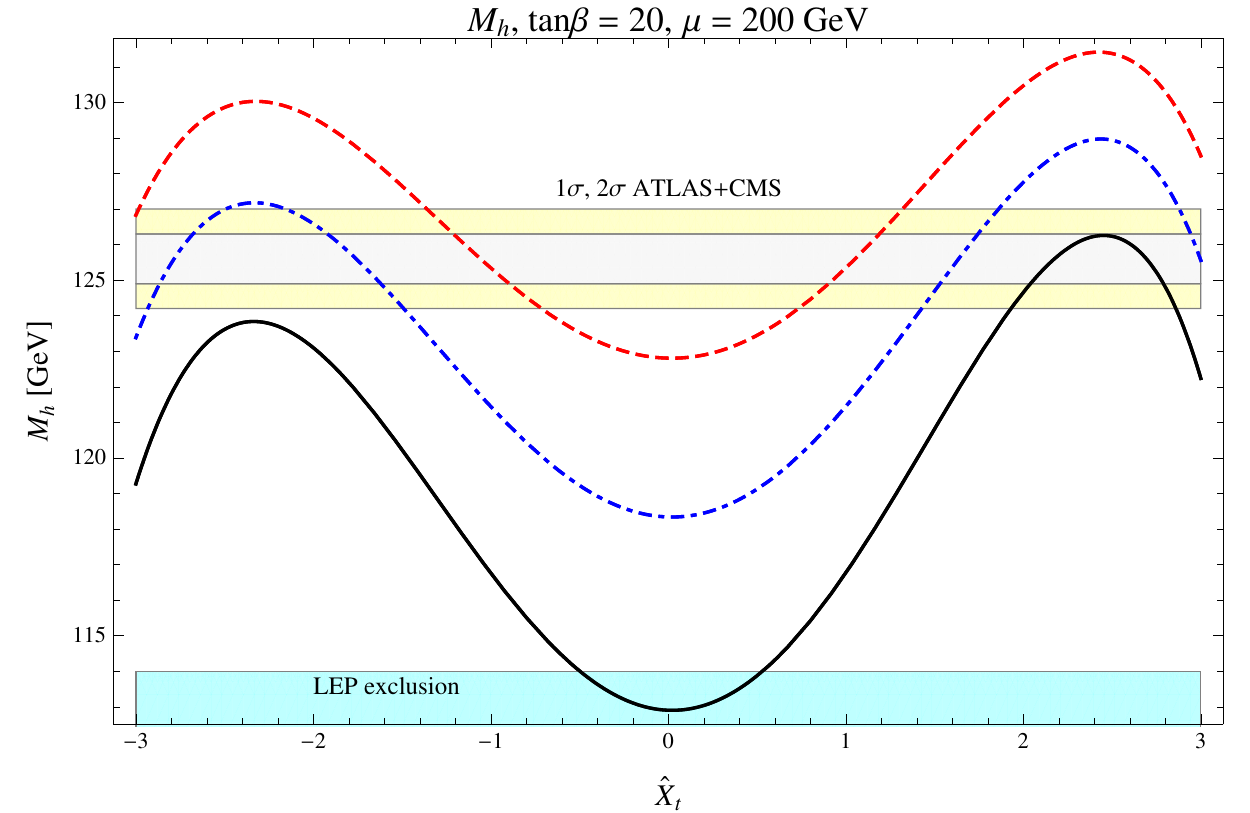}
\end{center}
\caption{Plot of Higgs mass $M_h$ vs. stop mixing parameter normalized by the SUSY scale, $\XtMS = X_t/M_S$. We have fixed the values $\tan\b = 20, \mu = 200$ GeV, and the (solid black, blue dot-dashed, red dashed) contours correspond to $M_S = (1, 2, 4)$ TeV.}
\label{fig:MhvXtMS}
\end{figure}

We can also plot the Higgs mass as a function of the normalized stop mixing parameter $\XtMS$, fixing the scale $M_S$, $\tan\beta$, and $\mu$. This is shown in Fig. \ref{fig:MhvXtMS}, where we have chosen $\tan\beta = 20, \mu = 200$ GeV, and plotted three curves for $M_S = 1, 2, 4$ TeV. The asymmetry in $\XtMS$, which was noted in \cite{Carena:2000dp} and \cite{Espinosa:2000df}, is due to the odd powers of $\XtMS$ in the $\mathcal{O}(\a_s \a_t)$ threshold correction to $\l_{\text{MSSM}} (M_S)$, Eq. (\ref{eqn:2loopasat}). For large $\tan\beta$ and $M_S = 1$ TeV, it is possible to obtain $M_h = 125.6$ GeV with $\XtMS > 0$ and near the maximal value. For $M_S = 2$ TeV, we require $|\XtMS| \sim 1.5$ TeV. We note that even for $M_S = 4$ TeV, $M_h = 125.6$ GeV is not achieved for zero mixing, which was also shown in the top-left plot of Fig. \ref{fig:tanbvMS}.

Lastly, we comment on some comparisons with existing calculations. We have generally presented Higgs masses which are lower than those computed by, e.g. \textsc{CPSuperH} \cite{Lee:2012wa}, \textsc{FeynHiggs} \cite{Heinemeyer:1998yj}, \textsc{SoftSUSY} \cite{Djouadi:2002ze}, \textsc{SPheno} \cite{Porod:2011nf}, and \textsc{H3M} \cite{Kant:2010tf} for $M_S \sim 1$ TeV. There are three differences between the calculations. First, we have used the NNLO value of $y_t$, which leads to a running top quark mass $\overline{m}_t (m_t)$ that is 2 GeV lower than the NLO value. Second, the electroweak running of $y_t$ has a large effect, since the $g_2^2$ contribution to $\beta_{y_t}^{(1)}$ is about $10\%$ that of the $g_3^2$ contribution. Since $y_t$ appears to the fourth power in both the one-loop $\b_\l^{(1)}$ and threshold corrections to $\l_{\text{MSSM}}$, these differences are significant. At higher scales, the running of $g_1, g_2$ in the tree-level $\l_{\text{MSSM}} (M_S)$ will also result in a lower Higgs mass. Together, these three effects can lead to disagreements of the order of a few GeV in $M_h$ from other approaches. We acknowledge that our calculation may still be missing important nonleading-log corrections.

%----------------------------------------------------------------------------------------
%	CONCLUSION
%----------------------------------------------------------------------------------------

\section{Conclusion} \label{sect:concl}
In this work we have presented three- and four-loop next-to-next-to-leading-log corrections to the lightest Higgs boson mass in the MSSM, in the approximation where the other MSSM scalars and gluino are heavy and controlled by a common scale $M_S$. We have compared the fixed-order result to the full resummation method for computing the Higgs mass and found that our four-loop formula with renormalization scale $Q=M_S$ is accurate up to scales of order a few tens of TeV. Using lower-loop truncations or the renormalization scale $Q=M_t$ leads to worse agreement with the more accurate resummed result. We also revisit a known accidental cancellation that appears in the three- and four-loop terms and conclude that partial three-loop results may overestimate the Higgs mass by a few GeV at large $M_S$ due to the absence of some of the cancelling terms. Our results include relevant corrections that were not present in previous calculations and become relevant when one computes the Higgs mass with greater precision at higher SUSY scales. In fact, even for $M_S \sim 1$~TeV, we find that these lower the Higgs mass by 2-4 GeV depending on the parameters of the stop sector. This has important implications for the definition of the soft supersymmetric breaking parameters in different SUSY scenarios.

\textbf{Note Added}

While we were finishing this work, Ref.~\cite{Hahn:2013ria} appeared that deals with similar issues in the diagrammatic approach. In their analysis, the three coupled SM RGE's for $y_t, g_3,$ and $\lambda$ were numerically integrated from $M_S$ to $m_t$, with the values $m_{\tilde{g}} = 1.6$ TeV, $\mu = M_2 = 1$ TeV, and $\tan\beta = 10$ for the MSSM parameters. Our results agree for values of $1 \leq M_S \leq 1.5$ TeV; however, in the case of maximal mixing, we do not reproduce the steep positive slope in the upper plot of Fig. 1 of \cite{Hahn:2013ria}. Further investigation is needed to resolve this discrepancy.

%----------------------------------------------------------------------------------------
%	ACKNOWLEDGEMENTS
%----------------------------------------------------------------------------------------

\begin{acknowledgments} \label{sect:acknow}
P.~D. and C.~W. would like to thank S.~Martin for a fruitful discussion. 
C.~W. would also like to thank P.~Kant, G.~Kane, H.~Rzehak, N.~Shah, and J.~Wells for useful discussions.
PD.~. is supported in part by U.S. Department of
Energy Grant No.~DE-FG02-04ER41286.
Work at ANL is supported in part by the U.S. Department of Energy under Contract No.~DE-AC02-06CH11357. 
G. L. acknowledges support from DOE Grant No.~DE-FG02-13ER41958.

\end{acknowledgments}

\appendix

%----------------------------------------------------------------------------------------
%	APPENDIX: SCHEME CONVERSIONS FOR COUPLINGS
%----------------------------------------------------------------------------------------

\section{Conversion between $\DRbar$ and $\MSbar$ Schemes} \label{app:schemes}

In this appendix, we present the conversions between the $\DRbar$ and $\MSbar$ schemes used in the literature for parameters appearing in the threshold corrections to $\l$. Notation used in this appendix is summarized in Table \ref{table:schemenotation}. We note that the one-loop $\a_t, \a_s$ conversions between the $\DRbar$ and the OS schemes are presented in \cite{Espinosa:2000df}.

From \cite{Espinosa:2000df}, we express the MSSM running top quark mass $\wt{m}_t$ in the $\DRbar$ scheme and the SM running top quark mass $\mb_t$ in the $\MSbar$ scheme in terms of the top quark pole mass $M_t$:
\ba{
\wt{m}_t^2 (Q) &= M_t^2 \Bigg\{ 1-  \frac83 \k \tilde{g}_3^2 \Big[5 + 3 \log\frac{Q^2}{\wt{m}_t^2} + \log\frac{\wt{M}_S^2}{Q^2} - \frac{\wt{X}_t}{\wt{M}_S} \Big] \\
& \ + \frac32 \k \wt{h}_t^2 \Big[ (1 + \cb^2) \Big(\frac12 - \log\frac{\wt{M}_S^2}{Q^2} \Big) + \sb^2 \Big( \frac83 + \log\frac{Q^2}{\wt{m}_t^2} \Big) 
- \muMS^2 f_2(\muMS) \Big) \Big] \Bigg\} , \nl
\mb_t^2 (Q) &= M_t^2 \Bigg\{ 1 - \frac83 \k g_3^2 \Big[ 4 + 3\log\frac{Q^2}{m_t^2} \Big] + \frac12 \k y_t^2 \Big[ 8 + 3\log\frac{Q^2}{m_t^2} \Big] \Bigg\} ,
}
% The difference arises from the treatment of the gauge boson contribution in the self-energy of the top quark. 
where
\be{
f_2(\muMS) = \frac1{1 - \muMS^2} \Big[1 + \frac {\muMS^2}{1 - \muMS^2} \log\muMS^2 \Big],
}
and the parameters in the one-loop corrections are actually scheme independent in our approximation, as any corrections would be of higher order. Using these two equations with $y_t = \tilde{h}_t \sb$, $\tilde{g}_3 = g_3$, and $\wt{X}_t/\wt{M}_S = X_t/M_S \equiv \XtMS$ to leading order, we can derive the relation between the top quark mass in the $\DRbar$ and $\MSbar$ schemes at $Q = M_S$:
\ba{
\wt{m}_t^2 (M_S) = \mb_t^2 (M_S) & \Bigg\{ 1 - \frac83 \k g_3^2 \Big[ 1 - \XtMS \Big]
+ \frac32 \k \frac{y_t^2}{\sb^2} \Big[ \frac12 (1 + \cb^2) - \muMS^2 f_2(\muMS) \Big] \Bigg\} . \label{eqn:mtDRMS}
}
To convert this expression into a relation between Yukawa couplings, we use
\ba{
\wt{m}_t = \tilde{h}_t \sb \frac{\tilde{v}}{\sqrt2}, \quad & \quad \mb_t = y_t \frac{v}{\sqrt2} , \\
v^2(Q) = \wt{v}^2(Q) &\Bigg\{ 1 + \frac12 \k \tilde{h}_t^2 \sb^2 \XtMS^2 \Bigg\} ,
}
so that
\ba{
\tilde{h}_t^2 (M_S) = \frac{y_t^2 (M_S)}{\sb^2} & \Bigg\{ 1- \frac83 \k g_3^2 \Big[ 1 - \XtMS \Big] 
+ \frac32 \k \frac{y_t^2}{\sb^2} \Big[ \frac12 (1 + \cb^2) + \frac13 \XtMS^2 \sb^2 - \muMS^2 f_2(\muMS) \Big) \Big] \Bigg\} .
}

\begin{table}[htb]
\begin{center}
\begin{tabular}{ccc}
Parameter & \quad $\DRbar$ \quad & \quad $\MSbar$, (w/, w/o) thresholds \quad \\
\hline
Top quark mass & $\wt{m}_t$ & $(m_t, \mb_t)$ \\
Stop mixing parameter & $\wt{X}_t$ & $(X_t, \overline{X}_t)$ \\
SUSY scale & $\wt{M}_S$ & $(M_S, \overline{M}_S)$ \\
Top quark Yukawa & $\tilde{h}_t$ & $(h_t , y_t)$ \\
$SU(3)_c$ gauge coupling & $\tilde{g}_3$ & $(g_3, g_3)$ \\
Higgs vev & $\tilde{v}$ & $(v, v)$
\end{tabular}
\end{center}
\caption{Notation for parameters in different schemes.}
\label{table:schemenotation}
\end{table}

The result for the top quark Yukawa at $Q = M_S$ can be checked with the expression found in \cite{Martin:2007pg}, for which we obtain, in Martin's notation,
\ba{
c_\l &= 12 \XtMS^2 \Big( 1 - \frac{\XtMS^2}{12} \Big) , \nl
c_{g_3} &= - \frac12 , \nl
c_{y_t} &= \frac43 (1 - \XtMS) , \\
c^\prime_{y_t} &= -\frac38 \frac{1 + \cb^2}{\sb^2} - \frac14 \XtMS^2 
+ \frac3{4\sb^2} \frac{\muMS^2}{1 - \muMS^2} \Big(1 + \frac {\muMS^2}{1 - \muMS^2} \log\muMS^2 \Big) , \nl
c_v &= \frac{\XtMS^2}4 . \notag 
}
These coefficients yield
\ba{
g_3 = \tilde{g}_3 & \Bigg\{ 1 + \frac12 \k \tilde{g}_3^2 \Bigg\} , \nl
y_t = \tilde{h}_t & \sb \Bigg\{ 1 + \frac43 \k \tilde{g}_3^2 (1 - \XtMS) - \k \tilde{h}_t^2 \Big[ \frac38 (1 + \cb^2) + \frac14 \XtMS^2 \sb^2 \\
& \quad - \frac34 \frac{\muMS^2}{1 - \muMS^2} \Big(1 + \frac {\muMS^2}{1 - \muMS^2} \log\muMS^2 \Big) \Big] \Bigg\} , \nl
v = \tilde{v} & \Bigg\{1 + \frac14 \k \tilde{h}_t^2 \sb^2 \XtMS^2 \Bigg\} . \notag
}
The correction to $g_3$ is less than $0.5\%$ across the range of $M_S$ considered in this paper, and is neglected since $g_3$ appears at two-loop order in $\l_{\text{MSSM}} (M_S)$. These agree with the previous relations.

The top quark mass at $Q = M_S$ used in Secs. 4 and 5 of \cite{Carena:2000dp} will be denoted $m_t (M_S)$, and it is related to $\mb_t(M_S)$ by
\be{
m_t (M_S) = \mb_t (M_S) \Bigg\{1 + \frac43 \k g_3^2 \XtMS \Bigg\} ,
}
i.e. $m_t (M_S)$ includes the one-loop term proportional to $g_3^2 \XtMS$ in Eq. (\ref{eqn:mtDRMS}). We have checked that the $g_3^2$ terms in these expressions agree with those of \cite{Carena:2000dp}. The top quark Yukawa coupling associated with this $m_t$ is $h_t\sb$, where $h_t$ is given in Eq. (\ref{eqn:hytMSbar}) in Sec. \ref{sect:matching}, with additional corrections. We can then write
\ba{
\tilde{h}_t^2 (M_S) = h_t^2 (M_S) & \Bigg\{ 1- \frac83 \k g_3^2
+ \frac32 \k \frac{y_t^2}{\sb^2} \Big[ \frac12 (1 + \cb^2) + \frac13 \XtMS^2 \sb^2 - \muMS^2 f_2(\muMS) \Big] \Bigg\} .
}

Let us now examine the MSSM parameters $M_S$ and $X_t$. If we include radiative corrections, the relation between on-shell stop masses $M_{\tt_i}$ and the $\DRbar$ running parameters $\wt{M}_S, \wt{X}_t, \wt{m}_t$ are
\ba{
M_{\tt_1}^2 = \wt{M}_S^2 - \wt{m}_t \wt{X}_t - \frac12 \text{Re} \Big[ \Pi_{\tt_L \tt_L} (M_{\tt_2}^2) + \Pi_{\tt_R \tt_R} (M_{\tt_2}^2) \Big] + \text{Re} \Pi_{\tt_L \tt_R} (M_{\tt_1}^2) , \\
M_{\tt_2}^2 = \wt{M}_S^2 + \wt{m}_t \wt{X}_t - \frac12 \text{Re} \Big[ \Pi_{\tt_L \tt_L} (M_{\tt_1}^2) + \Pi_{\tt_R \tt_R} (M_{\tt_1}^2) \Big] - \text{Re} \Pi_{\tt_L \tt_R} (M_{\tt_1}^2) .
}
The self-energies $\Pi$, which can be found in \cite{Espinosa:2000df, Pierce:1996zz, Donini:1995wh}, contain one-loop corrections, and are the same in both $\MSbar$ and $\DRbar$ schemes. From this, we find
\ba{
\wt{M}_S^2 = M_S^2 \Bigg\{ 1 & - \frac{m_t^2}{M_S^2} \Big(1 - \frac{\wt{m}_t^2}{m_t^2} \Big) \Bigg\} , \\
\wt{X}_t = X_t \frac{m_t}{\wt{m}_t} , & \\
\frac{\wt{X}_t}{\wt{M}_S} = \frac{X_t}{M_S} \Bigg\{ 1 &+ \frac43 \k g_3^2 \Big(1 + \frac{m_t^2}{M_S^2} \Big) \nl
& + \frac32 \k h_t^2 \Big[ \frac12 + \frac{\cb^2}2 - \muMS^2 f_2(\muMS) \Big) \Big] \Big(1 + \frac{m_t^2}{M_S^2} \Big) \Bigg\} . \label{eqn:XtDRMSMSSM}
}
We will ignore the $m_t^2/M_S^2$ corrections here, as these terms appear at two-loop order. If we elect to use the SM $\MSbar$ top quark mass $\mb_t$ in lieu of $m_t$, then $\overline{X}_t, \overline{M}_S$ are obtained from $\wt{X}_t, \wt{M}_S$ using the above equations and replacing $m_t \rightarrow \mb_t$ where it appears:
\ba{
\frac{\wt{X}_t}{\wt{M}_S} = \frac{\overline{X}_t}{\overline{M}_S} \Bigg\{ 1 & 
+ \frac43 \k g_3^2 \Big(1 - \frac{\overline{X}_t}{\overline{M}_S} \Big) \Big(1 + \frac{\mb_t^2}{\overline{M}_S^2} \Big) \nl
& + \frac32 \k \frac{y_t^2}{\sb^2} \Big[ \frac12 + \frac{\cb^2}2 - \muMS^2 f_2(\muMS) \Big] \Big(1 + \frac{\mb_t^2}{\overline{M}_S^2} \Big) \Bigg\} . \label{eqn:XtDRMS}
}

%----------------------------------------------------------------------------------------
%	APPENDIX: BETA-FUNCTIONS
%----------------------------------------------------------------------------------------

\section{$\b$ Functions for the Fixed-Order Computation} \label{app:betafuncs}

The two-loop SM $\b$ functions were first computed in \cite{Machacek:1983tz, Machacek:1983fi, Machacek:1984zw}; we have used the equations from Appendix A of \cite{Arason:1991ic}, with corrections to $\b_\l^{(2)}$ in \cite{Luo:2002ey}. $\b_\l^{(1)} \big|_{\tilde{\chi}}$, the one-loop electroweakino contribution to $\b_\l^{(1)}$ can be found in Appendix C of \cite{Haber:1993an}, and together with $\b_{y_t}^{(1)} \big|_{\tilde{\chi}}$ also in in \cite{Giudice:2011cg}.
The two-loop $\b$ functions with just $\l, g_3, y_t$ can also be found in \cite{Martin:2007pg}. $\b_\l^{(3)}$ was computed in \cite{Harlander:2008ju, Kant:2010tf, Chetyrkin:2013wya}. We use the expressions for the three-loop $\b$ functions for $g_3, y_t, \l$ from \cite{Buttazzo:2013uya}, which also contains references to their computations.

When comparing the $\beta$ functions in these references, one must be careful of conventions for $\l$ and $v$. We have adopted those of \cite{Arason:1991ic}, with a Higgs potential of the form 
\be{
V(\F) = -\frac{m^2}2 |\F|^2 + \frac{\l}2 |\F|^4 ,
}
and a Higgs doublet in the broken phase of the form
\be{
\F = \binom{0}{\frac{v + h}{\sqrt2}} , \quad v \sim 246 \GeV.
}
For the case $m_A \sim M_S$, we include here the SM $\MSbar$ $\b$ functions for $\l, g_3, y_t$ used in the fixed-order computation for performing the RG running between $Q = M_t$ and $Q = M_S$. We include $y_b, y_\tau, g_2,$ and $g_1$ only in $\b_X^{(1)}$, for $X = g_3, \l$. We have assumed that the electroweak couplings do not run. The one-loop electroweakino contribution to $\b_\l$ is denoted by $\b^{(1)}_\l \big|_{\tilde{\c}}$, and will be multiplied by a different logarithmic enhancement, namely $L_\m = \log(M_S/\m) = \log(M_S/M_{1,2})$.

\ba{
% \b_\l^{(1)}
\b^{(1)}_\l &= 12 \l^2 + 4 \l ( 3y_t^2 + 3y_b^2 + y_\t^2) - 4 ( 3y_t^4 + 3y_b^4 + y_\t^4 ) \nl
& \quad - 9 \l \Big( g_2^2 + \frac15 g_1^2 \Big) + \frac94 \Big( g_2^4 + \frac25 g_1^2 g_2^2 + \frac3{25} g_1^4 \Big) , 
\label{eqn:betalambda1} \\
\b^{(1)}_\l \big|_{\tilde{\c}} &= \Bigg[ 6\l \Big( g_2^2 + \frac15 g_1^2 \Big) - \Big( g_2^2 + \frac35 g_1^2 \Big)^2 - 4g_2^4 (1 - 2 \sb^2 \cb^2) \Bigg] \frac{L_\m}L , 
\label{eqn:betalambda1chi} \\
% \b_\l^{(1,1)}
\b_\l^{(1,1)} &= \dt{\l} \cdot \Bigg[ 12 (2\l + y_t^2) - 9 \Big( g_2^2 + \frac13 g_1^2 \Big) + 6 \Big( g_2^2 + \frac15 g_1^2 \Big) \frac{L_\m}{L} \Bigg] + y_t \dt{y_t} \cdot 24 (\l - 2y_t^2) ,
\label{eqn:betalambda11} \\
% \b_\l^{(1,2)}
\b_\l^{(1,2)} &= \dnt{\l}{2} \cdot \Bigg[ 12 (2\l + y_t^2) - 9 \Big( g_2^2 + \frac13 g_1^2 \Big) + 6 \Big( g_2^2 + \frac15 g_1^2 \Big) \frac{L_\m}{L} \Bigg] + \Big( \dt{\l} \Big)^2 \cdot 24 \nl
& \quad + \dt{\l} \Big( y_t \dt{y_t} \Big) \cdot 48 + y_t \dnt{y_t}{2} \cdot 24 (\l - 2y_t^2) + \Big( \dt{y_t} \Big)^2 \cdot 24 (\l - 6 y_t^2) , 
\label{eqn:betalambda12}
}
\ba{
% \b_\l^{(1,3)}
\b_\l^{(1,3)} &= \dnt{\l}{3} \cdot \Bigg[ 12 (2\l + y_t^2) - 9 \Big( g_2^2 + \frac13 g_1^2 \Big) + 6 \Big( g_2^2 + \frac15 g_1^2 \Big) \frac{L_\m}{L} \Bigg] \nl
& \quad + \dnt{\l}{2} \cdot 72 \Bigg[ \dt{\l} + y_t \dt{y_t} \Bigg]
+ \dt{\l} \cdot 72 \Bigg[ \Big( \dt{y_t} \Big)^2 + y_t \dnt{y_t}{2} \Bigg] \nl
& \quad + y_t \dnt{y_t}{3} \cdot 24 (\l - 2y_t^2) + \dnt{y_t}{2} \dt{y_t} \cdot 72 (\l - 6y_t^2) - y_t \Big( \dt{y_t} \Big)^3 \cdot 288 , 
\label{eqn:betalambda13} \\
% \b_\l^{(1,4)}
\b_\l^{(1,4)} &= \dnt{\l}{4} \cdot \Bigg[ 12 (2\l + y_t^2) - 9 \Big( g_2^2 + \frac13 g_1^2 \Big) + 6 \Big( g_2^2 + \frac15 g_1^2 \Big) \frac{L_\m}{L} \Bigg] 
+ \dnt{\l}{3} \cdot 96 \Bigg[ \dt{\l} + y_t \dt{y_t} \Bigg] \nl
& \quad + \dnt{\l}{2} \cdot 72 \Bigg[ \dnt{\l}{2} + 2 \Bigg[ \Big( \dt{y_t} \Big)^2 + y_t \dnt{y_t}{2} \Bigg] \Bigg] 
+ \dt{\l} \cdot 96 \Bigg[ 3 \dnt{y_t}{2} \dt{y_t} + y_t \dnt{y_t}{3} \Bigg] \nl
& \quad + y_t \dnt{y_t}{4} \cdot 24 (\l - 2y_t^2) + \dnt{y_t}{3} \dt{y_t} \cdot 96 (\l - 6y_t^2) + \Big( \dnt{y_t}{2} \Big)^2 \cdot 72 (\l - 6y_t^2) \nl
& \quad - y_t \dnt{y_t}{2} \Big( \dt{y_t} \Big)^2 \cdot 1728 - \Big( \dt{y_t} \Big)^4 \cdot 288 .
\label{eqn:betalambda14}
}
\ba{
% \b_\l^{(2)}
\b_\l^{(2)} &= -78 \l^3 - 72 \l^2 y_t^2 + 80 \l g_3^2 y_t^2 - 3 \l y_t^4 - 64g_3^2 y_t^4  + 60 y_t^6 , 
\label{eqn:betalambda2} \\
% \b_\l^{(2,1)}
\b_\l^{(2,1)} &= \dt{\l} \cdot (-234 \l^2 - 144 \l y_t^2 + 80g_3^2 y_t^2 - 3 y_t^4) + g_3 \dt{g_3} \cdot 32 y_t^2 (5 \l - 4 y_t^2) \nl
& \quad + y_t \dt{y_t} \cdot 4 (-36 \l^2 + 40 \l g_3^2 - 3 \l y_t^2 - 64 g_3^2 y_t^2 + 90 y_t^4) , 
\label{eqn:betalambda21} \\
% \b_\l^{(2,2)}
\b_\l^{(2,2)} &= \dnt{\l}{2} \cdot (-234 \l^2 - 144 \l y_t^2 + 80 g_3^2 y_t^2 - 3 y_t^4) \nl
& \quad + \dt{\l} \Bigg[ \dt{\l} \cdot (-36) (13 \l + 4 y_t^2 ) + g_3 \dt{g_3} \cdot 320 y_t^2 + y_t \dt{y_t} \cdot 8(-72\l + 40 g_3^2 - 3 y_t^2) \Bigg] \nl
& \quad + g_3 \dnt{g_3}{2} \cdot 32y_t^2 (5 \l - 4 y_t^2) + \Big( \dt{g_3} \Big)^2 \cdot 32 y_t^2 (5 \l - 4 y_t^2) \nl
& \quad + \Big( g_3 \dt{g_3} \Big) \Big( y_t \dt{y_t} \Big)\cdot 128 (5 \l -8 y_t^2) \nl
& \quad + y_t \dnt{y_t}{2} \cdot 4 (-36 \l^2 + 40 \l g_3^2 - 3 \l y_t^2 - 64 g_3^2 y_t^2 + 90 y_t^4) \nl
& \quad + \Big( \dt{y_t} \Big)^2 \cdot 4 (-36 \l^2 + 40 \l g_3^2 - 9 \l y_t^2 - 192 g_3^2 y_t^2 + 450 y_t^4) .
\label{eqn:betalambda22}
}
\ba{
% \b_\l^{(3)}
\b_\l^{(3)} &= \frac{\l^3}2 \Bigg( 6011.35 \frac{\l}2 + 873 y_t^2 \Bigg) + \l^2 y_t^2 (1768.26 y_t^2 + 160.77 g_3^2) \nl
& \quad + 2 \l y_t^2 (-223.382 y_t^4 - 662.866 g_3^2 y_t^2 + 356.968 g_3^4) \nl
& \quad + 4y_t^4 (-243.149 y_t^4 + 250.494 g_3^2 y_t^2 - 50.201 g_3^4) ,
\label{eqn:betalambda3}
}
\ba{
% \b_\l^{(3,1)}
\b_\l^{(3,1)} &= \dt{\l} \Bigg[ 6011.35 \l^3 + \frac32 \cdot 873 \l^2 y_t^2 + 2\l y_t^2 (1768.26 y_t^2 + 160.77 g_3^2) \nl
& \quad \quad + 2y_t^2 (-223.382 y_t^4 - 662.866 y_t^2 g_3^2 + 356.968 g_3^4) \Bigg] \nl
& \quad + y_t \dt{y_t} \Bigg[ 873 \l^3 + 2\l^2 (2 \cdot 1768.26 y_t^2 + 160.77 g_3^2) \nl
& \qquad \qquad + 4\l (3 \cdot (-223.382) y_t^4 + 2 \cdot (-662.866) y_t^2 g_3^2 + 356.968 g_3^4) \nl
& \qquad \qquad + 8y_t^2 (4 \cdot (-243.149) y_t^4 + 3 \cdot 250.494 y_t^2 g_3^2 + 8 \cdot (-50.201) g_3^4) \Bigg] \nl
& \quad + g_3 \dt{g_3} \Bigg[ 2 \cdot 160.77 \l^2 y_t^2 + 4\l y_t^2 ( (-662.866) y_t^2 + 2 \cdot 356.968 g_3^2) \nl
& \qquad \qquad + 8y_t^4 ( 250.494 y_t^2 + 2 \cdot (-50.201) g_3^2) \Bigg] .
\label{eqn:betalambda31}
}

The above couplings are evaluated at the scale $M_S$, and we use the following $\beta$-functions to evolve $g_3, y_t$ from $M_S$ down to $M_t$.
\ba{
% \b_{g_3}}^{(1)}
\b_{g_3}^{(1)} &= -g_3^3 \Bigg[ 11 - \frac23 \Nfl \Bigg] , 
\label{eqn:betag31} \\
% \b_{g_3}}^{(1,1)}
\b_{g_3}^{(1,1)} &= - \dt{g_3} \cdot 3g_3^2 \Bigg[ 11 - \frac23 \Nfl \Bigg] , 
\label{eqn:betag311} \\
%\b_{g_3}^{(1,2)} &= -\Bigg[ g_3^2 \dnt{g_3}{2} + 2 g_3 \Big( \dt{g_3} \Big)^2 \Bigg] \cdot 3 \Big( 11 - \frac23 \Nfl \Big) .
% \b_{g_3}}^{(2)}
\b_{g_3}^{(2)} &= -g_3^3 \Bigg[ \Big( 102 - \frac{38}3 \Nfl \Big) g_3^2 + 2 y_t^2 \Bigg] .
\label{eqn:betag32} 
%\b_{g_3}^{(2,1)} &= -\dt{g_3} \cdot g_3^2 \Bigg[ 5 \Big(102 - \frac{38}3 \Nfl \Big) g_3^2 + 6 y_t^2 \Bigg] - y_t \dt{y_t} \cdot 4 g_3^3 .
}
\ba{
% \b_{y_t}}^{(1)}
\b_{y_t}^{(1)} &= y_t \Bigg[ \frac92 y_t^2 + \frac32 y_b^2 + y_\tau^2 - 8 g_3^2 - \frac94 g_2^2 - \frac{17}{20} g_1^2 \Bigg] , 
\label{eqn:betayt1} \\
\b_{y_t}^{(1)} \big|_{\tilde{\chi}} &= \Bigg[ \frac32 y_t \Big( g_2^2 + \frac15 g_1^2 \Big) \Bigg] \frac{L_\m}{L} , 
\label{eqn:betayt1chi} \\
% \b_{y_t}}^{(1,1)}
\b_{y_t}^{(1,1)} &= - g_3 \dt{g_3} \cdot 16 y_t 
+ \dt{y_t} \cdot \Bigg[ \frac{27}2 y_t^2 - 8g_3^2 - \frac94 g_2^2 - \frac{17}{20} g_1^2 + \Big( g_2^2 + \frac15 g_1^2 \Big) \frac{L_\m}{L} \Bigg] ,
\label{eqn:betayt11} \\
% \b_{y_t}}^{(1,2)}
\b_{y_t}^{(1,2)} &= -  \Bigg[ g_3 \dnt{g_3}{2} + \Big( \dt{g_3} \Big)^2 \Bigg] \cdot 16 y_t - g_3 \dt{g_3} \dt{y_t} \cdot 32 \nl
& \quad + \dnt{y_t}{2} \cdot \Bigg[ \frac{27}2 y_t^2 - 8g_3^2 - \frac94 g_2^2 - \frac{17}{20} g_1^2 + \Big( g_2^2 + \frac15 g_1^2 \Big) \frac{L_\m}{L} \Bigg] 
+ \Big( \dt{y_t} \Big)^2 \cdot 27 y_t ,
\label{eqn:betayt12}
}
\ba{
% \b_{y_t}}^{(1,3)}
\b_{y_t}^{(1,3)} &= - \Bigg[ g_3 \dnt{g_3}{3} + 3 \dnt{g_3}{2} \dt{g_3} \Bigg] \cdot 16y_t 
- \Bigg[ g_3 \dnt{g_3}{2} \dt{y_t} + \Big( \dt{g_3} \Big)^2 \dt{y_t} + g_3 \dt{g_3} \dnt{y_t}{2} \Bigg] \cdot 48 \nl
& \quad + \dnt{y_t}{3} \Bigg[ \frac{27}2 y_t^2 - 8g_3^2 - \frac94 g_2^2 - \frac{17}{20} g_1^2 + \Big( g_2^2 + \frac15 g_1^2 \Big) \frac{L_\m}{L} \Bigg] \nl
& \quad + y_t \dnt{y_t}{2} \dt{y_t} \cdot 81 + \Big( \dt{y_t} \Big)^3 \cdot 27 .
\label{eqn:betayt13}
}
\ba{
% \b_{y_t}}^{(2)}
\b_{y_t}^{(2)} &= y_t \Bigg[ \frac32 \l^2 - 6\l y_t^2 - \Big( \frac{404}3 - \frac{40}9 \Nfl \Big) g_3^4 + 36 g_3^2 y_t^2 - 12y_t^4 \Bigg] , 
\label{eqn:betayt2} \\
% \b_{y_t}}^{(2,1)}
\b_{y_t}^{(2,1)} &= \dt{\l} \cdot 3y_t (\l - 2 y_t^2) + g_3 \dt{g_3} \cdot 4y_t \Big[ - \Big( \frac{404}3 - \frac{40}9 \Nfl \Big) g_3^2 + 18 y_t^2 \Big] \nl
& \quad + \dt{y_t} \cdot \Bigg[ \frac32 \l^2 - 18 \l y_t^2 - \Big( \frac{404}3 - \frac{40}9 \Nfl \Big) g_3^4 + 108 g_3^2 y_t^2 - 60 y_t^4 \Bigg] .
\label{eqn:betayt21}
}
\ba{
% \b_{y_t}}^{(3)}
\b_{y_t}^{(3)} = y_t \Bigg[ & -\frac92 \l^3 + \frac{15}{16} \l^2 y_t^2 + \l y_t^2 (99 y_t^2 + 8 g_3^2) \nl
& + 58.6028 y_t^6 - 157 y_t^4 g_3^2 + 363.764 y_t^2 g_3^4 - 619.35 g_3^6 \Bigg] .
\label{eqn:betayt3}
}
We set the number of active quark flavors $\Nfl = 6$ for running above the scale $M_t$. Note that $\l(M_t)$ appears in $\b_{y_t}^{(2)}, \b_{y_t}^{(2,1)}, \b_{y_t}^{(3)}$. We approximate it using the tree-level MSSM value, Eq. (\ref{eqn:lambdatree}), in $\b_{y_t}^{(2,1)}, \b_{y_t}^{(3)}$, and an effective one-loop value in $\b_{y_t}^{(2)}$ that also includes the one-loop stop thresholds and one-loop running with $\b_{\l}^{(1)}$, with all parameters evaluated at $Q = M_t$.

%----------------------------------------------------------------------------------------
%	BIBLIOGRAPHY
%----------------------------------------------------------------------------------------

%----------------------------------------------------------------------------------------

\end{document}